\documentclass[fleqn,usenatbib]{mnras}

\usepackage{newtxtext,newtxmath}
\usepackage[T1]{fontenc}

\DeclareRobustCommand{\VAN}[3]{#2}
\let\VANthebibliography\thebibliography
\def\thebibliography{\DeclareRobustCommand{\VAN}[3]{##3}\VANthebibliography}

\usepackage{graphicx}	% Including figure files
\usepackage{amsmath}	% Advanced maths commands

\usepackage{multicol}
\usepackage[dvipsnames]{xcolor}
\usepackage[normalem]{ulem}

\newcommand{\ergs}{erg~s$^{-1}$}
\newcommand{\minusone}{$^{-1}$}
\newcommand{\minustwo}{$^{-2}$}
\newcommand{\minusthree}{$^{-3}$}

\newcommand{\pdot}{$\dot{P}$}

\newcommand{\plm}{$\pm$}
\newcommand{\frb}{FRB~20200120E}

\newcommand{\psr}{PSR~B1821$-$24A}
\newcommand{\psrj}{J1824$-$2452A}
\newcommand{\psrgeyer}{PSR~J0540$-$6919}

\newcommand{\mus}{$\mu$s}

\newcommand{\digifil}{\verb|digifil}
\newcommand{\psrchive}{\verb|PSRCHIVE}
\newcommand{\dspsr}{\verb|dspsr}
\newcommand{\pazi}{\verb|pazi}
\newcommand{\pdmp}{\verb|pdmp}
\newcommand{\pam}{\verb|pam}

\newcommand{\fil}{\verb|.fil}

\newcommand{\ar}{\verb|.ar}

\newcommand{\dmunits}{pc~cm\minusthree}
\newcommand{\tscatt}{$\tau_{\text{scatt}}$}

\usepackage[flushleft]{threeparttable}

\title[Searching for links between MSPs and FRBs]{Searching for links between energetic millisecond pulsars and repeating fast radio bursts} 

\author[R. J. van Ruiten et al.]{
R. J. van Ruiten,$^{1}$ %\thanks{E-mail: }% (KTS)} % and others
J. W. T. Hessels,$^{1,2,3,4}$
S. Bhandari,$^{1,2,5,6}$
P. Chawla,$^{1,2}$
A. Gopinath,$^{1}$
D. M. Hewitt,$^{1}$
\newauthor
\hspace{0.001mm} 
K. Nimmo$^{7}$
and M. P. Snelders$^{1,2}$
\\
$^{1}$Anton Pannekoek Institute for Astronomy, University of Amsterdam, Science Park 904, 1098 XH, Amsterdam, The Netherlands\\
$^{2}$ASTRON, Netherlands Institute for Radio Astronomy, Oude Hoogeveensedĳk 4, 7991 PD Dwingeloo, The Netherlands\\
$^{3}$Department of Physics, McGill University, 3600 rue University, Montr\'eal, QC H3A~2T8, Canada\\
$^{4}$Trottier Space Institute, McGill University, 3550 rue University, Montr\'eal, QC H3A 2A7, Canada\\
$^{5}$Joint institute for VLBI ERIC, Oude Hoogeveensedĳk 4, 7991 PD Dwingeloo, The Netherlands\\
$^{6}$CSIRO Space and Astronomy, Australia Telescope National Facility, PO Box 76, Epping, NSW 1710, Australia\\
$^{7}$MIT Kavli Institute for Astrophysics and Space Research, Massachusetts Institute of Technology, 77 Massachusetts Ave, Cambridge, MA 02139, USA\\
}

\date{Accepted XXX. Received YYY; in original form ZZZ}

\pubyear{2026}

\begin{document}
\label{firstpage}
\pagerange{\pageref{firstpage}--\pageref{lastpage}}
\maketitle

\begin{abstract}
The unexpected localization of the repeating \frb\ to a globular cluster challenges conventional FRB models based on magnetars formed via core collapse. One alternative model suggests that \frb\ is a millisecond pulsar (MSP) producing giant pulses (GPs). 
To test this hypothesis, we compared the characteristics of \frb\ bursts with the GPs of the most energetic Galactic MSP known, M28A (\psr), using observations with the Parkes (Murriyang) telescope's Ultra Wideband Low-frequency (UWL) receiver. 
Our analysis provides insight into the spectral structure and frequency extent of M28A's GPs, revealing broad-band spectra spanning $700-4000$\,MHz (in some cases) with complex spectral peaks. 
We find that known M28A GP characteristics persist at this bandwidth, such as durations, luminosities, periodicity, wait-time, and energy distribution. 
A sub-band search for narrow-band GPs yielded no detection of genuinely narrow-band GPs. However, we do find narrow-band spectral peaks of $\sim 100$\,MHz bandwidth, a similar scale observed for \frb's bursts.
Compared to \frb's bursts, M28A's GPs have $50 \times$ shorter durations, $10^5 \times$ lower spectral luminosities, clear periodicity (vs. no periodicity), a purely Poissonian wait-time distribution (vs. quasi-Poissonian), and generally broad-band spectra with narrow-band peaks (vs. only narrow-band bursts). Both sources show a steep energy distribution and minor dispersion measure variability. 
Our study finds no strong links between M28A and \frb. However, we cannot rule out the possibility that \frb\ is a rare and unique type of MSP with no Galactic analogue. Furthermore, higher-cadence monitoring of M28A, for hundreds to thousands of hours, might reveal rare but extremely luminous pulses.

% {\color{blue}
% \begin{itemize}
%     \item briefly describe the aims, methods, and main results of the paper
%     \item single paragraph not more than 250 words
%     \item no references should appear in the abstract
% \end{itemize}
% }
\end{abstract}

\begin{keywords}
fast radio bursts -- radio continuum: transients -- pulsars: individual: B1821$-$24A % -- FRB -- %, J1824−2452A
\end{keywords}

%%%%%%%%%%%%%%%%%%%%%%%%%%%%%%%%%%%%%%%%%%%%%%%%%%

%%%%%%%%%%%%%%%%% BODY OF PAPER %%%%%%%%%%%%%%%%%%
\section{Introduction}
Fast radio bursts (FRBs) are extremely bright (fluences of $\sim 0.01-1000$\,Jy~ms), short-duration (\mus$~-~$s), and often tremendously distant (Mpc$~-~$Gpc), extragalactic radio pulses \citep{cordes_2019_araa,petroff_2019_aarv, petroff_2022_aarv}. A small fraction ($\sim 3\%$) of FRBs observed thus far show repeat bursts \citep{spitler_2016_natur, fonseca_2020_apjl, chime_2023_apj}. Their emission has been observed over a broad frequency band, ranging from $110$\,MHz \citep{pleunis_2021_apjl} to $8$\,GHz \citep[][actually $\sim10$\,GHz in the source reference frame]{gajjar_2018_apj}. Repeaters often show time-frequency drifts in their bursts (also colloquially referred to as the `sad-trombone effect'), suggestive of a radius-to-frequency mapping effect \citep[e.g.,][]{lyutikov_2020_apj}, as well as a narrow fractional bandwidth of $10-30$\% \citep[e.g.,][]{hessels_2019_apjl}. 

The number of known FRBs and those with known host galaxies is increasing rapidly. However, the physical origins of FRBs continue to be debated \citep[see][for a catalogue of proposed theories]{platts_2019_phr}. While observations strongly suggest that at least some FRBs originate from magnetars -- young neutron stars (NSs) powered by the decay of their exceptionally large magnetic field \citep{bochenek_2020_natur, chime_2020_natur_galacticfrb} -- the diversity of FRB properties and locations suggests that this is not the full explanation.

Recently, the repeating \frb\ was pinpointed to a globular cluster (GC) in the M81 galactic system \citep{kirsten_2022_natur, bhardwaj_2021_apjl}. This location is surprising, as GCs typically host ancient stellar populations, in contrast to the young core-collapse magnetars invoked in promising FRB models \citep{metzger_2017_apj, margalit_2018_mnras}. Furthermore, \frb\ is remarkably nearby ($3.6$\,Mpc), placing it a few times closer than any other known extragalactic FRB host \citep{bhardwaj_2021_apjl}. \citet{kirsten_2022_natur} propose alternative scenarios to explain the unexpected location of \frb. They suggest that \frb\ could be a young magnetar formed through unconventional mechanisms, such as accretion-induced collapse (AIC) of a white dwarf (WD) or merger-induced collapse (MIC) of binary star systems. These binary systems can consist of WD-WD, NS-WD, or NS-NS pairs, all of which are relatively common in GCs \citep{kremer_2025}. Of these scenarios, the merger of a WD-WD binary is considered the most plausible, as it is a common phenomenon in the central cores of GCs \citep{kremer_2021_apj, kremer_2021_apjl}.

Another distinct hypothesis proposed by \citet{kirsten_2022_natur} is that \frb\ is a recycled millisecond pulsar (MSP) producing giant pulses (GPs). GCs are one of the most efficient MSP producers, with a $10^{4} - 10^{5}\times$ higher formation rate per unit stellar mass compared to the Galactic field \citep[][and reference therein]{hessels_2015_aska}.
\citet{cordes_2016_mnras} discussed the possibility of a GP origin of some FRBs, based on Crab GPs. The Crab pulsar produces the most energetic GPs of all known Galactic pulsars. The highest inferred brightness
temperature for a Crab `nanoshot' is $T_\text{B} \sim 2\times10^{41}$\,K \citep[excluding relativistic effects;][]{hankins_2007_apj}, comparable to that observed for \frb\ bursts \citep[$T_\text{B} \sim 3\times10^{41}$\,K;][]{nimmo_2022_natas}. This comparison is relevant because both MSPs and young pulsars like the Crab can exhibit similarly strong magnetic fields at the light cylinder $-$ a parameter thought to be critical for GP generation \citep{romani_2001_apjl, wang_2019_scpma} $-$ achieved via either rapid rotation (MSPs) or strong surface magnetic fields (Crab).
\citet{connor_2016_mnras} hypothesize that a Crab-like pulsar could emit super-GPs sporadically bright enough to be detectable at a few hundred Mpc. 

There is an ongoing debate on the origin of \frb. For example, both \citet{lu_2022_mnras} and \citet{kremer_2021_apjl} support the idea by \citet{kirsten_2022_natur} that the most likely scenario is that \frb\ is a magnetar formed via MIC of a WD binary. However, both have a different view on the possibility of an MSP origin. \citet{lu_2022_mnras} argue against an MSP origin, based on higher magnetic field requirement estimates compared to typical MSPs. On the other hand, \citet{kremer_2021_apjl} do not rule out an MSP origin. They estimate that a typical recycled MSP could reproduce \frb's time-averaged luminosity, taking into account radio emission efficiencies. Furthermore, from a rates perspective, they suggest that MSPs could account for the density of repeating FRBs inferred from \frb, but only if their duty cycle for producing bursts similar to the M81 FRB is small. 

\citet{nimmo_2022_natas} show that, compared to other known repeaters \citep{pleunis_2021_apj}, the bursts from \frb\ have similarly high ($\sim 100\%$) linear and low ($\lesssim 10\%$) circular polarization fractions, along with a modest Faraday rotation measure (RM) of $-29.8$\,rad~m\minustwo. However, compared to FRBs in general, they also have atypically short durations ($50-100$\,\mus); $100\times$ lower isotropic-equivalent spectral luminosities ($L\sim10^{28}$\,erg~s\minusone~Hz\minusone); a steep energy distribution (power-law index of $\alpha \sim 2.39$); and a stable dispersion measure (DM) with variations of $\Delta\text{DM} \lesssim 0.15$\,\dmunits\ over $>10$\,months \citep{nimmo_2023_mnras}.

The exceptionally narrow temporal profiles of the bursts from \frb\ and their uncommonly low luminosities -- the bright radio burst from the Galactic magnetar SGR 1935+2154 \citep[sometimes called `FRB~200428';][]{chime_2020_natur_galacticfrb, bochenek_2020_natur} has even a $10\times$ higher spectral luminosity than the extragalactic \frb\ bursts --  are an additional line of evidence supporting a different physical origin compared to other repeaters. Furthermore, the bursts also show $\sim 100$ nanosecond and $\sim 1$ microsecond temporal substructure \citep{nimmo_2022_natas, majid_2021_apjl} that is similar to that seen for GPs \citep{knight_2006_apj, hankins_2016_apj}. 
However, it is worth noting that \frb\ is not the only FRB with such temporal substructure; recent studies identified other FRBs showing microsecond substructure \citep{nimmo_2021_natas}, isolated microsecond duration bursts \citep{snelders_2023_natas}, and clustered forests of microshots during wider bursts \citep{hewitt_2023_mnras}.

Multi-wavelength observations are another important means to constrain the physical nature of \frb. Deep, simultaneous radio and X-ray observations of \frb\ have failed to detect prompt X-ray emission at the times of radio bursts, or any persistent X-ray emission \citep{pearlman_2025}. Those authors find that \frb\ is unlikely to be associated with ultraluminous X-ray bursts, magnetar-like giant flares, or an SGR~1935+2154-like intermediate flare, but they cannot conclusively rule out the magnetar hypothesis.

To test the hypothesis that \frb\ is an MSP emitting GPs, it is essential to understand the general characteristics of GPs. GPs are intense radio pulses emitted by some rotation-powered radio pulsars, significantly ($>10-30\times$) brighter than their average pulses. GP-emitting sources are rare, with only approximately 16 pulsars known to emit GPs \citep[see, e.g.,][and references therein]{kazantsev_2018_arxiv}. GPs are further distinguishable by three properties seen in some, but not necessarily all, sources: i) a power-law energy distribution, unlike the normal pulsar emission (main pulse) that follows a log-normal or a Gaussian distribution \citep{burkespolaor_2012_mnras}; ii) narrow intrinsic widths typically less than a few microseconds \citep{hankins_2007_apj}; and iii) sporadic occurrence within specific, narrow spin phase windows. 
While some GPs are identified as broad-band ($>0.8$\,GHz) through dual-band observations \citep[e.g.,][]{sallmen_1999_apj}, a few GPs are detected as narrow-band ($\lesssim 200$\,MHz), e.g. by \citet[][]{geyer_2021_mnras, thulasiram_2021_mnras, bij_2021_apj}. The latter is important, as narrow-band GPs serve as a link between pulsars and repeating FRBs. GPs can also be ultra-short in duration; \citet{hankins_2007_apj} reported the observation of Crab GPs with nanoshot structures as narrow as $0.4$\,ns. Such brief durations indicate extremely high brightness temperatures, implying that the GPs originate from non-thermal and coherent emission processes \citep{hankins_2003_natur}. The exact emission mechanism, however, remains unknown. 
In the case of MSPs, GPs typically occur at the trailing edge of the average radio profile components and are often coincident with pulsed X-ray emission \citep{romani_2001_apjl, cusumano_2003_aa, knight_2006_apj, bilous_2015_apj}. The link with X-ray emission has led to suggestions that GPs may originate from the same magnetospheric region as high-energy emission, possibly representing radio components of that emission. However, the small sample size of GP-emitting MSPs limits definitive conclusions \citep{abbate_2020_mnras}, and we would not expect such emission to be detectable at the distance of \frb\ \citep{pearlman_2025}. 

In this study, we search for links between energetic MSPs and repeating FRBs by testing the hypothesis that \frb's bursts are GPs from an MSP. Currently observed GPs are five orders-of-magnitude less luminous than the lowest-observed FRB luminosities, see Figure~3 in \citet{nimmo_2022_natas} which motivates us to study the most energetic MSP in the Galaxy. We thus selected M28A (\psr, also called \psrj), an isolated MSP ($P_{\text{spin}}=3.054$\,ms) located in the GC Messier 28 (M28) \citep{lyne_1987_natur}, which lies at a distance of $d\text{\textsubscript{M28A}} = 5.6 \pm 0.3\,\text{kpc}$ \citep{oliveira_2022_aa}. M28A is a well-known GP emitter \citep{romani_2001_apjl, knight_2006_apj, bilous_2015_apj}. \citet{knight_2006_apj} measured GP durations as short as $20-100$\,ns up to less than a few microseconds, and measured microsecond and nanosecond-structure using a time-resolution of $7.8125$\,ns at $2.7$ and $3.5$\,GHz.

M28A is an extraordinary MSP in many aspects, and is an outlier compared to the bulk of the MSP population. It has an uncommonly large spin-down of \pdot\ $= 1.62 \times 10^{-18}$\,s s\minusone\ \citep{foster_1988_apjl, verbiest_2009_mnras}, which is about two orders of magnitude larger than typical $\dot{P}$ values for MSPs. Moreover, it is one of the youngest among the old recycled accretion/rotation powered MSPs, with an estimated spin-inferred characteristic age of about $2.99 \times 10^{7}$\,yr \citep{cognard_2004_apjl}. It has the second-largest magnetic field strength at the light cylinder among MSPs of $B_{\text{LC}}\sim 7.2 \times 10^5$\,G (where $B_{\text{LC}} \propto P^{-2.5}\dot{P}^{0.5}$), and the highest surface magnetic field strength of all known MSPs of $B_{\text{surf}}\sim 2.25 \times 10^9$\,G (where $B_{\text{surf}} \propto [P \dot{P}]^{0.5}$ and $B_{\text{surf}} = B_{\text{LC}} [c/\Omega R]^{3}$, with $c$ being the speed of light, $\Omega$ the angular velocity, and R the radius of the NS).  

Furthermore, M28A has the highest-known MSP spin-down luminosity of $\dot{E} \sim 2.2 \times 10^{36}$\,\ergs\ (where $\dot{E} \propto P $\minusthree$ \dot{P}$), making it the most energetic MSP known. This increases our chances of detecting anomalously bright GPs, which could link them to FRBs by bridging the observed energy gap between MSPs and FRBs, see Figure~3 in \citet{nimmo_2022_natas}.

M28A emits electromagnetic radiation at multiple frequencies. It emits in radio \citep{lyne_1987_natur}, X-ray \citep{danner_1994_apjl, rots_1998_apj}, and is a bright $\gamma$-ray emitter \citep{abdo_2013_apjs, wu_2013_apjl, johnson_2013_apj} -- the latter further illustrating the extraordinary nature of this MSP. Attempts to link the multi-wavelength components remain inconclusive, suggesting a complex relationship between emission regions \citep{johnson_2013_apj}. The trailing edges of the two X-ray components coincide with the two rotational phases of the GPs, and one of the pulsed $\gamma$-ray components is aligned with the rotational phase of the GPs \citep[see, e.g., Figure~1~(b) and (c) in][]{bilous_2015_apj}.

Previous observations of M28A's GPs are band-limited (observed bandwidths up to $0.8$\,GHz), and therefore lack information about the spectra over larger instantaneous bandwidths. Our new study presents, for the first time, an instantaneous $3.3$-GHz broad-band analysis of M28A's GPs with $2$-\mus\ time resolution using the 64-m Murriyang telescope at the Parkes Observatory.

Section~\hyperref[sec:obs]{2} describes the observational setup. Section~\hyperref[sec:methods]{3} describes the methodology and the analysis techniques. Section~\hyperref[sec:results]{4} gives an overview of the results. In Section~\hyperref[sec:discussion]{5}, we discuss the potential links between the energetic Milky Way MSP M28A and the M81 repeater \frb. In Section~\hyperref[sec:conclusions]{6} we conclude and discuss future research ideas and improvements to the methodology.

\section{Observations}
\label{sec:obs}
We performed observations using the Ultra Wideband Low-frequency (UWL) receiver of the 64-m Murriyang telescope at the Parkes Observatory in New South Wales, Australia. The UWL receiver is able to measure an instantaneous $3.3$-GHz bandwidth spanning $704-4032$\,MHz. The signal pre-processor digitizes the data and generates $26$ sub-band data streams, each with a $128$-MHz bandwidth. These sub-band data streams are processed separately by the signal processor system backend called `Medusa', using Graphics Processing Units (GPUs). The system temperature is approximately $22$\,K for $\sim60$\% of the band. It increases near the band edges. From $1000$\,MHz down to $704$\,MHz the system temperature rises steeply, reaching $\sim40$\,K. From $3400$\,MHz to $4000$\,MHz, it increases to $\sim30$\,K \citep[see Figure 4 in][]{hobbs_2020_pasa}.

We recorded all four polarisation products -- AA$^*$, BB$^*$, Re[A$^*$B] and Im[A$^*$B] --where A and B represent the two orthogonal polarisation signals in a linear basis with complex sampling, and the $^*$ symbol denotes the complex conjugate. The data were collected using real-time coherent dedispersion for a DM of $119.9$\,\dmunits, with an exceptionally high time resolution of $2$\,\mus, and a frequency resolution of $4$\,MHz. This setup produced data rates of approximately $1$\,TB per hour. The observations (PI: van Ruiten; Project Code: P1151) were conducted on April 7, April 12, and June 7, 2022, referred to throughout the paper as sessions O1, O2, and O3, respectively. Refer to Table~\ref{tab:obs_results} for further observation details.

\section{Methods and analysis}
\label{sec:methods}
\subsection{Average radio profile}
We calculated an average radio profile using a set of software tools from the Pulsar Comprehensive High-speed Analysis and Archive Toolkit (\psrchive|, \citet{hotan_2004_pasa})\footnote{\url{https://psrchive.sourceforge.net/}}: \dspsr| (see, e.g., \citet{vanstraten_2011_pasa}), \pazi|, \pdmp|, and \pam|. 

We used \digifil| (tool in \dspsr|) to combine the two polarisation products AA$^*$ and BB$^*$ from the raw data (provided in PSRFITS search mode format, \verb|.sf|) into total-intensity Stokes~I (I = $\langle\text{AA}^*\rangle$ +  $\langle\text{BB}^*\rangle$), stored in filterbank (\fil|) files, to reduce the data volume and accelerate the folding process. We folded the data using \dspsr| and an M28A rotational ephemeris from \citet{manchester_2005_aj}\footnote{\url{https://www.atnf.csiro.au/research/pulsar/psrcat/}}, for which we updated the DM to $\text{DM}(\text{MJD} 59737)=119.906$\,\dmunits. Each folded profile was stored in an archive-file (\ar|-file).

We processed the \fil|-files by dedispersing (between channels) and folding them with \dspsr|, setting the integration time parameter to one second. Each second of data was segmented into intervals matching the rotational period, and the segments were summed for each frequency channel, producing a 1-s folded dynamic spectrum with Stokes~I intensity for all frequencies per phase bin. The folded profiles were then summed to create the average folded profile over an integration time of one and a half hours. 

This duration was more than sufficient to achieve high signal-to-noise ratio (S/N). To reduce data size and processing time for generating the average profile, we used $512$ phase bins, with each bin corresponding to a time resolution of $P\text{\textsubscript{M28A}} / \text{N\textsubscript{phase bins}} = 3054.3257~\text{\mus} / 512=5.965~\text{\mus}$ or $0.0019$ spin periods, roughly three times the native time resolution of the data. This slight downsampling in time was done to mitigate noise but avoid washing out profile features. We preserved the original frequency resolution of $4$\,MHz using $832$ frequency channels.

We used \pazi| to manually flag channels to remove radio frequency interference (RFI). Using \pdmp|, both time-phase and frequency-phase profiles were fitted to obtain S/N-optimized linear best-fit estimates for DM and spin period. 

\subsection{Pipeline for GP detection}
To avoid bias, we searched for GPs occurring at any rotational phase, even though previous studies have found M28A's GPs to be clustered at specific rotational phases \citep{bilous_2015_apj}.
To detect GPs, we adapted and modified an existing single-pulse search pipeline \citep{snelders_2023_natas}. We used the open-source software \verb|PRESTO| for most of the data processing tasks, including the initial RFI excision, dedispersion, and single-pulse search \citep{ransom_2011_ascl}. To assist in candidate classification, we used \verb|FETCH|, a machine-learning model based on dynamic spectrum images and DM-time diagrams \citep{agarwal_2020_mnras}.

Just as with the average profile determination, we converted the raw \verb|.sf| files into \verb|.fil| files for the single-pulse search using \verb|digifil|, summing polarizations to Stokes~I. Unlike for the average pulse profile, for the individual GPs we preserved the native recorded time resolution of $2$\,\mus.

For the single-pulse search, we used \texttt{PRESTO}'s \verb|single_pulse_search.py|, which applies boxcar matched filtering. We set a minimum boxcar width of 1 sample, i.e. $2$\,\mus, and a maximum boxcar width of $300 \times t_{\text{samp}}$ ($=600$\,\mus) -- both with an integrated S/N threshold of $5.0$. While this increased the number of low-S/N GP detections, it also raised the false positive rate.

To address multiple detections of the same event at neighbouring trial DMs, we used the \verb|Single-pulse Searcher| \citep[\texttt{SpS.py},][]{michilli_2018_mnras}, which grouped neighbouring events together. To further mitigate false positives, we rejected grouped candidates with excessively large widths ($>100$\,\mus) and/or low S/N ($<7.0$). Among the detected astrophysical candidates, only events with widths near $2$\,\mus\ (within a factor two) were identified, and no rare bursts with widths on the order of 10s of microseconds were observed.

For further classification, we used \verb|FETCH| to assign a probability to each candidate, indicating whether it is astrophysical or not. A \verb|.h5| file was created for each candidate, compatible with \verb|FETCH|, using the \verb|your_candmaker.py| script from Your Unified Reader (\verb|your|).\footnote{\url{https://github.com/thepetabyteproject/your}} We used \verb|FETCH|’s pre-trained model ‘a’ and set a probability threshold of $p > 0.5$ to minimise false positives while retaining genuine GPs. To validate the model’s applicability to our dataset, we visually inspected all candidates with $p > 0.5$, which were dominated by pulses with widths of $2-4$\,\mus\, and typically high ($p > 0.9$) probabilities. All of these were confirmed to be genuine pulses. We also checked a subset of lower-probability candidates (including some with widths $>4$\,\mus), none of which appeared astrophysical. This visual verification gave us confidence that the pre-trained model performed reliably for our data and pulse characteristics.

To manage the large data volume efficiently, we automated the GP search process. The resulting pipeline included five automated steps: 1) conversion of raw \verb|.sf| files to \verb|.fil| files, 2) organising data file structures to ensure smooth integration across all pipeline components, 3) application of \verb|PRESTO| tools for identifying single-pulse candidates, 4) application of \verb|YOUR| and \verb|FETCH| tools for candidate classification, and 5) extraction of GPs from the raw data.

The automated pipeline ensured a systematic and efficient approach to GP detection in the extensive dataset.

\subsection{Sub-band search}
Narrow-band radio bursts have reduced S/N when averaged across the whole band as opposed to a tailored sub-band where the burst is emitting. Consequently, narrow-band bursts might not appear as single peaks in the total-intensity time-series of a broad observing bandwidth and would therefore not be detected in single-pulse searches \citep[see, e.g., Figure 3 in][showing the detection of an extremely narrow-band repeat burst from FRB~20190711A with the Parkes UWL receiver]{kumar_2021_mnras}. 

Therefore, to investigate the potential presence of isolated narrow-band GPs in the full $3.3$-GHz band, we conducted a sub-banded search. For the first two observations, we divided the full band into three sub-bands, each spanning $1664$\,MHz, overlapping by half their bandwidths. For the third observation, we conducted a more thorough sub-band search, using $56$~sub-bands (corresponding to $3 \times 1664$\,MHz, $7 \times 832$\,MHz, $15 \times 416$\,MHz, and $31 \times 208$\,MHz sub-bands, each of the four sets overlapping by half their bandwidths). Due to its computational expense, we applied the detailed sub-band search only to observation O3, which had the most GP detections in the full-band search, thus increasing the chances of detecting potential narrow-band bursts.

We used a S/N-threshold of $7.0$ for the full-band and $1664$\,MHz sub-band detections, and $9.0$ for the smaller sub-bands, since all of low-S/N candidates in the smaller-sized sub-bands were false positives, confirmed by manual examination of the dynamic spectra.

\subsection{GP characterization}
\label{sec:gp_characterization}
In our search for connections between the energetic Galactic MSP M28A and the M81-repeater \frb, we examined various characteristics of M28A's GPs. These characteristics can be compared with the observed burst properties of the M81-repeater, as reported by \citet{nimmo_2023_mnras}.
The key burst properties we analysed for comparison are durations, isotropic spectral luminosities, energy distribution, wait-time distribution, DM variability, spectral structure and spectral indices, decorrelation bandwidths, and scattering timescales.

M28A's GPs are known to be emitted in two distinct rotational phase windows. \citet{bilous_2015_apj} split their GP sample from M28A based on phase and did not find any significant differences in observed characteristics between the two samples. Therefore, in our work we treat the Parkes M28A GP sample as a whole.

\subsubsection{Durations}
We investigated both observed burst durations and estimated intrinsic burst durations. The single-pulse search provided observed width estimates based on the total intensity profile across the ultra-wide $3.3$-GHz frequency range, extending down to $704$\,MHz. Examination of the pulses at the highest frequencies -- where scattering is minimal -- suggests that none of our GPs are temporally resolved, given the 2-$\mu$s sampling of our observations. Despite a small offset between the DM used for real-time coherent dedispersion ($119.9$\,\dmunits) and the DM determined post-observation, with offsets of $\Delta \text{DM} = 0.015, 0.014,~\text{and}~0.006$\,\dmunits\ for O1, O2, and O3, respectively, there is no significant intra-channel DM smearing compared to the time resolution of the data.

\subsubsection{DM variability}
The single-pulse search determines the S/N-maximized DM for each GP, which can lead to over- or under-dedispersion. To determine a more precise value for the DM and DM variability, we used the \verb|pdmp| script on 1-minute segments of 10 minutes of folded data. Note, however, that a strong scattering tail potentially complicates a robust DM determination.

\subsubsection{Isotropic spectral luminosities}
The isotropic-equivalent GP luminosity is determined as $L_{\nu,iso}=4\pi d^2 S_{\nu\text{,peak}}$, where $d$ is the distance to the source  
and $S_{\nu\text{,peak}}$ is the peak flux density of the pulse calculated using the measured peak S/N and the radiometer equation \citep{cordes_2003_apj}. Fluences are calculated by integrating $S_{\nu}$ over the pulse width and full bandwidth.

We assume a $20-30\%$ uncertainty in the flux density, primarily due to fluctuations in the system temperature. We used a single system temperature of $22$\,K, reflecting the average value for $60$\% of the band, which coincides with the region occupied by most of the GPs. The single-pulse search provided DM-optimized S/N, which occasionally resulted in over-dedispersion due to the scattering tail at lower frequencies. Therefore, we reprocessed the data for a sample of bright GPs, applying a single DM estimated with \verb|pdmp| for dedispersion and normalizing the bandpass. This approach allowed us to obtain better estimates of peak-S/N values and therefore better estimates of luminosities. 

\subsubsection{S/N distribution}
We used DM-optimized peak S/N values from the single-pulse search to determine the cumulative S/N distribution of the GPs, regardless of the rotational phase window of occurrence. We fitted the distribution with a power-law model using the method of least-squares, defining the power-law indices $\alpha$ through the relation $R \propto \text{S/N}^{-\alpha}$, where $R$ is the rate of bursts with signal-to-noise $> \text{S/N}$. 

\subsubsection{Wait-time distribution}
Wait times, defined as the time differences between consecutive GPs, provide insight into burst clustering and potentially also emission mechanisms. We conducted a wait-time analysis for each individual observation to investigate their distribution and GP rates.
The cumulative wait-time distributions are fitted with a Poissonian cumulative distribution function (CDF) defined as   
\begin{equation}
    P_{\text{Poisson}} = 1 - e^{-t_{\text{wait}}\mathcal{R}_{\text{P}}}, 
    \label{eq:poissondistr}
\end{equation}
where $t_\text{wait}$ is the wait time and $\mathcal{R}_{\text{P}}$ the Poissonian rate.

\subsubsection{Spectra and spectral indices}
We obtained the GP spectra by summing across the burst width in a frequency-dependent way to account for the scattering tail. To isolate the relevant region in frequency-time space, we used an average of bright GPs aligned at $t=0$ to define where GPs are expected. This selected region is divided into sub-bands, where the time integration width increases toward lower frequencies to reflect the increased scattering. Each sub-band is normalized to the one corresponding to the most prominent part of the scattering tail. After normalization, each sub-band is averaged over time, and the resulting values are combined to form a 1D frequency spectrum. This process is illustrated for the brightest GPs in Figure~\ref{fig:family_plot_18GPs_4col}.

Depending on the type of analysis, we applied a Savitzky-Golay filter \citep{savitzky_1964_anach} to smooth the spectra while preserving their structure. For spectral index measurements and visualisation, we used a filter with window length $wl = 8$ and polynomial order $po = 1$. For the frequency spectra shown in the side panels of the dynamic spectra (see Figure~\ref{fig:family_plot_18GPs_4col}), we used stronger smoothing with $wl = 16$ and $po = 2$ to enhance clarity. For the autocorrelation function (ACF) analysis discussed below, no filtering was applied. The Savitzky-Golay filter settings were determined by trial-and-error, and were chosen to enhance visibility of the spectral structure over noise fluctuations and RFI-zapping, while not over-smoothing.

\subsubsection{ACF analysis}
We determine the decorrelation bandwidth of the few brightest GP spectra using an autocorrelation function (ACF) analysis. This provides a measure of the characteristic frequency structure of the bursts. Under certain conditions $-$ such as sufficient frequency resolution and a single dominant scattering screen $-$ this observable can be interpreted as an estimate of the scintillation bandwidth, which relates to scattering (see Section~\ref{sec:tscatt}). 
However, the observed frequency structure could also arise from intrinsic emission processes or local propagation effects, and is not necessarily due to interstellar scintillation alone.

We compute one-dimensional ACFs of the frequency spectra using the expression defined by \citet{marcote_2020_natur}:
\begin{equation}
    \text{ACF}(\Delta \nu) = \frac{ \sum\limits_{i}(S(\nu_i)-\bar{S})(S(\nu_i+\Delta \nu)-\bar{S}) }{\sqrt{ \sum\limits_{i}(S(\nu_i)-\bar{S})^2 } \sqrt{ \sum\limits_{i} (S(\nu_i+\Delta \nu)-\bar{S})^2}},
    \label{eq:acf}
\end{equation}
where $\nu$ is the frequency, $\Delta \nu$ the frequency lag, and $\bar{S}$ the mean of the spectrum. 
Only valid (non-flagged) data -- i.e., excluding channels affected by RFI or band-edge artifacts -- are included in the summation, and the zero-lag term is excluded from the computation.

The decorrelation bandwidth is defined as the half-width at half-maximum (HWHM) of the central peak in the ACF \citep{rickett_1990_araa}. 
The HWHM is obtained by least-squares fitting the central part of the ACF around zero frequency lag with a Lorentzian function $f_{\text{L}}$, defined as
\begin{equation}
    f_{\text{L}}(\mu=0) = \frac{A}{\pi}\frac{\Delta \nu_{\text{HWHM}}}{\nu^2 +\Delta \nu_{\text{HWHM}}^2}, 
\end{equation}
where $A$ is the amplitude, $\Delta \nu_{\text{HWHM}}$ the decorrelation bandwidth, $\nu$ the frequency lag and where the Lorentzian centre $\mu$ is set to zero lag.

\subsubsection{Scattering timescales}
\label{sec:tscatt}

With adequate frequency resolution, the observed decorrelation bandwidth can be interpreted as the scintillation bandwidth, $\Delta \nu_{\text{scint}}$, assuming that interstellar scattering is the dominant process.

Assuming a single scattering screen and Kolmogorov turbulence, the scattering timescale $\tau_{\text{scatt}}$ can be used to determine the scintillation bandwidth using the relation $\Delta \nu_{\text{scint}} \sim 1/(2\pi \tau_{\text{scatt}})$ \citep{lorimer_2004_hpa}.

To determine the scattering timescales, we modelled the total-intensity profile of a GP as the convolution of a Gaussian distribution (representing the fast rise) with a one-sided exponential decay. This model is expressed as:
\begin{equation}
    f_{\text{pulse, ts}}(t, \tau_{\text{scatt}}, \mu, \sigma, \mathcal{A}) = \mathcal{A} \times (\mathcal{N}(\mu, \sigma^2) * e^{-(t - \mu)/\tau_{\text{scatt}}}), 
\label{eq:pulse_profile_model_ts}
\end{equation}
where $t$ is time, $\tau_{\text{scatt}}$ is the scattering timescale, $\mathcal{N}$ is a normal probability density function centered around time $\mu$ with variance $\sigma^2$, and $\mathcal{A}$ is a scaling constant. In our GP time-series, $\mu$ approximately matches the time of arrival of the pulse, centered at $t=0$.

We fitted this function to GP total-intensity profiles, each divided into a set of $200$-MHz sub-bands that overlap by half, covering central frequencies up to around $2100$\,MHz, beyond which scattering effects become comparable to the time resolution. We determined the scattering frequency dependence (power-law index $\beta$) by fitting a power law to the best-fit scattering times as a function of frequency, where $\tau \propto \nu^{-\beta}$. The power-law index $\beta$ was then compared to the expected value for a single, thin, infinitely wide screen with Kolmogorov turbulence ($\beta=4.4$). Deviations from this value might indicate a screen that does not extend infinitely, or possibly multiple scattering screens \citep{geyer_2016_mnras}.

At a reference frequency of $1$\,GHz, which is also used in other studies, we determined $\tau_{\text{scatt}}$ for the three brightest GPs and used $\sim 1/(2\pi \tau_{\text{scatt}})$ to predict the scintillation bandwidth. We compared our measured scattering and scintillation values with predictions from the NE2001 model, which estimates insterstellar medium (ISM) scattering timescales based on the distance or DM of the source \citep{cordes_2002_arxiv}. 

\section{Results}
\label{sec:results}
Here we present the results of the analyses described in \S\ref{sec:methods}. Table\,\ref{tab:obs_results} includes the number of bursts detected in each observation and the average rate per observation.

\subsection{Average pulse profile}
\begin{figure}
	\includegraphics[width=\columnwidth]{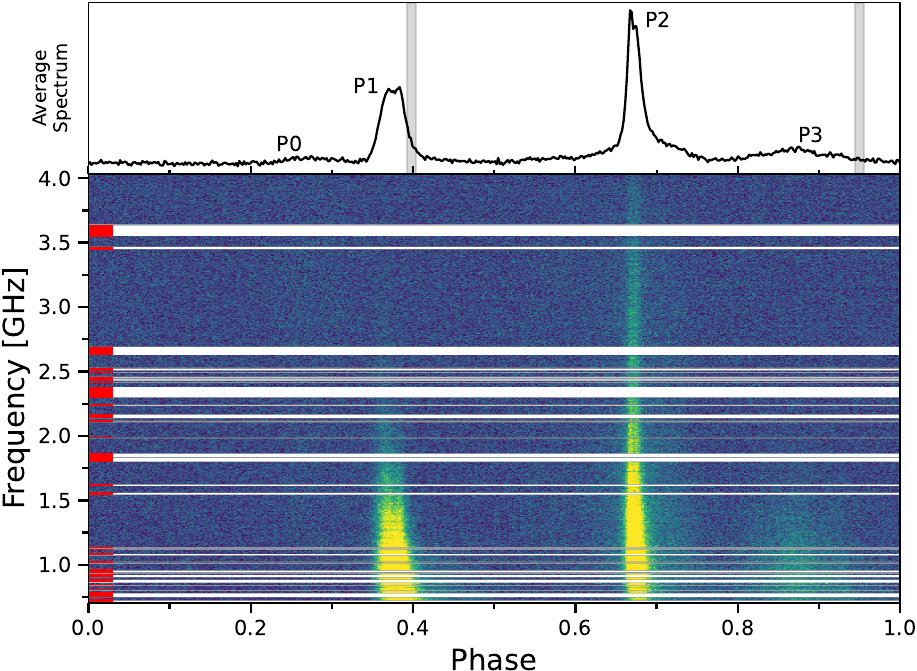}
    \caption{\textit{Lower panel:} Average dynamic spectrum of M28A (Stokes~I, observed with the UWL-receiver in the $0.7-4.0$\,GHz range) with a frequency resolution of $4$\,MHz and coherent dedispersion. Each rotational phase bin corresponds to $5.965$\,\mus\ ($\sim 3\times t_{\text{samp}}$). Folding is done over $1.$6\,hr in O1, using the rotational ephemeris from \citet{manchester_2005_aj}, with an updated DM of $\text{DM}(\text{MJD 59737})=119.906$\,\dmunits. Red bars indicate flagged RFI channels. \textit{Upper panel:} Frequency- and time-averaged pulse profile of M28A with components labeled ($\text{P0}-\text{P3}$). M28A's GPs fall into two narrow phase windows at the trailing edges of P1 and P3, indicated by the shaded grey regions, which represent the phase ranges reported by \citet{bilous_2015_apj}.} 
    \label{fig:average_profile_M28A}
\end{figure}

Figure~\ref{fig:average_profile_M28A} shows the average pulse profile of M28A, which exhibits four main components, P0 to P3. The wide-band ($3.3$-GHz) observations show that P2 extends the most, on average, towards higher frequencies, spanning the entire band. \citet{knight_2006_apj} and \citet{bilous_2015_apj} showed that M28A's GPs fall into two distinct, narrow rotational phase windows, corresponding to the trailing edges of both pulsed X-ray components. The first phase window also corresponds to the trailing edge of the radio component P1. The second phase-window corresponds to the trailing edge of the P3 radio component -- more specifically, with the secondary bump at the right side of the main P3 peak. \citet{bilous_2015_apj} used their data to determine GP characteristics such as wait-time distribution and energy distribution for both phases separately, and did not find significant differences between these samples. Therefore, we combined all GPs in our data for analysis, regardless of their arrival phase. 

In total, $156$ GPs exceeding the S/N threshold of $7.0$ were detected during $8.77$ hours of observations using Parkes' UWL receiver, see Table~\ref{tab:obs_results}. The average GP rate is comparable between observations.

\begin{table}
    \caption{M28A Parkes observation details, Project Code: P1151. For all observations, the frequency range is $0.704 - 4.32$\,GHz, and the frequency/time resolution is $4.0$\,MHz / $2.0$\,\mus. The dates are topocentric, the durations correspond to the time used in this research, and the number of GPs is for the detections above the S/N threshold of 7.0.}
    \label{tab:obs_results}
    \centering
    \small 
    \setlength{\tabcolsep}{10pt}
    \begin{threeparttable}
        \begin{tabular}{ c c c c c c }
            \hline
            \hline
            ID & Start Obs. & Duration & N$_{\text{GPs}}$ & Avg. Rate \\
                        & \text{[MJD]}        & [hr]               &                            & [hr\minusone]  \\
            \hline
            O1      & 59676.712           & 2.15               & 38                        & 17.7 \\ 
            O2      & 59681.711           & 3.13               & 55                        & 17.6 \\ 
            O3      & 59737.526           & 3.49               & 63                        & 18.1 \\ 
            \hline
        \end{tabular}
    \end{threeparttable}
\end{table}

\subsection{Pulse wait times}
\begin{figure}
	\includegraphics[width=\columnwidth]{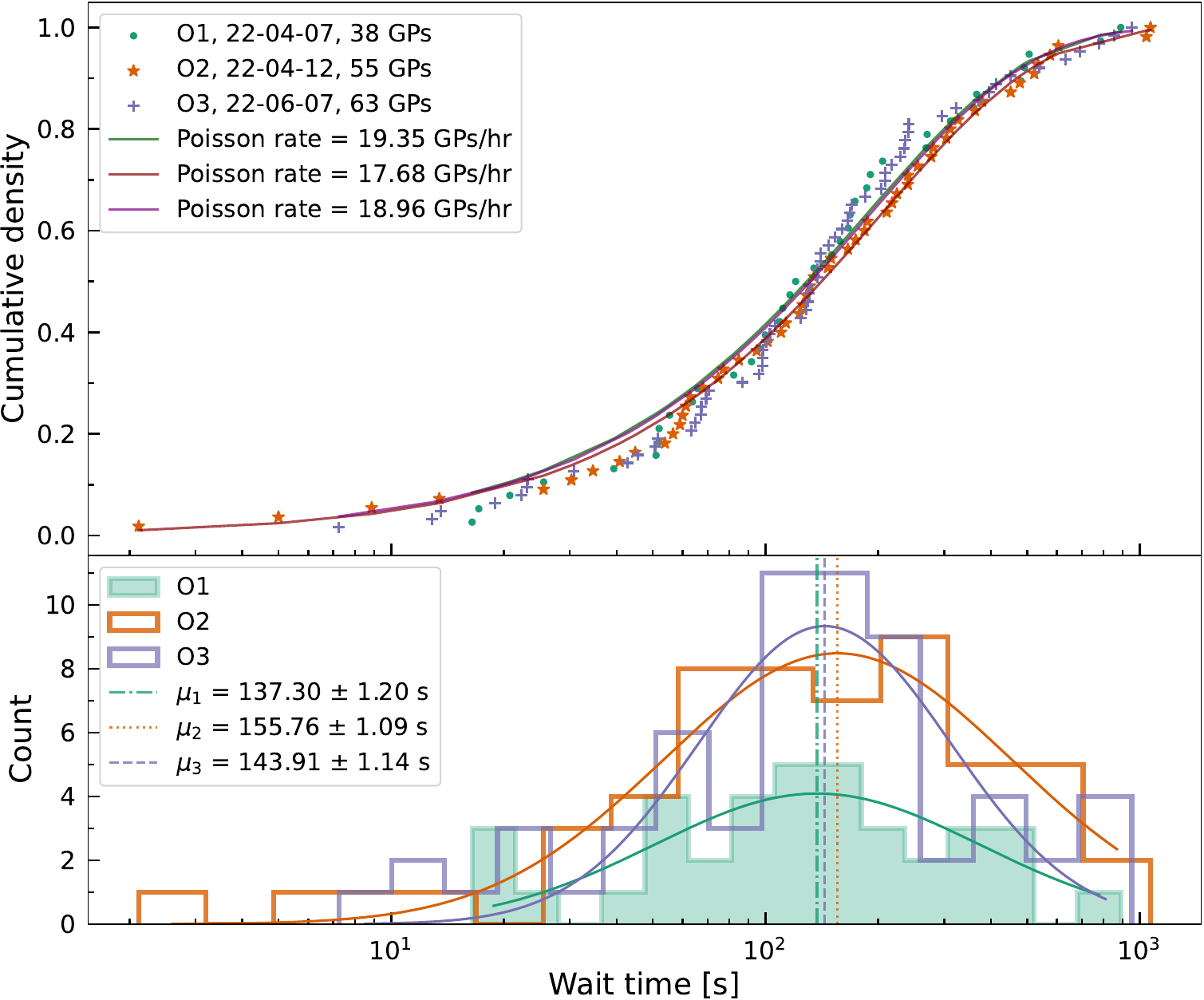}
    \caption{\textit{Lower panel:} Wait-time distributions of the GPs detected in O1, O2 and O3 (green, orange and purple, respectively). A single log-normal is fitted to each distribution using least-squares. Best-fit log-normal mean wait-times: $\mu = 137.3$\,s for O1, $\mu = 155.7$\,s for O2 and $\mu = 143.9$\,s for O3 (horizontal dashed lines). \textit{Upper panel:} Corresponding cumulative distributions with best-fit Poisson rates: $\mathcal{R}_{\text{P}} = 19.4$\,hr\minusone\ for O1, $\mathcal{R}_{\text{P}} = 17.7$\,hr\minusone\ for O2 and $\mathcal{R}_{\text{P}} = 19.0$\,hr\minusone\ for O3.}
    \label{fig:waittime_distr}
\end{figure}

In each Parkes observation, the GP wait-times follow a log-normal distribution, indicating a mean wait-time of approximately 150\,s, see Figure~\ref{fig:waittime_distr}. The cumulative distributions were fitted with a Poisson cumulative distribution function (CDF), resulting in the best-fit GP rate parameters of approximately $\mathcal{R}_{\text{P}} = 19$\,hr\minusone, $\mathcal{R}_{\text{P}} = 18$\,hr\minusone, and $\mathcal{R}_{\text{P}} = 19$\,hr\minusone\ for O1, O2, and O3, respectively. The corresponding mean wait-times, assuming a log-normal fit, were approximately $\mu = 137.30 \pm 1.20$\,s, $\mu = 155.76 \pm 1.09$\,s, and $\mu = 143.91 \pm 1.14$\,s, using the least-squares fitting method.

\subsection{S/N distribution}
\begin{figure}
	\includegraphics[width=\columnwidth]{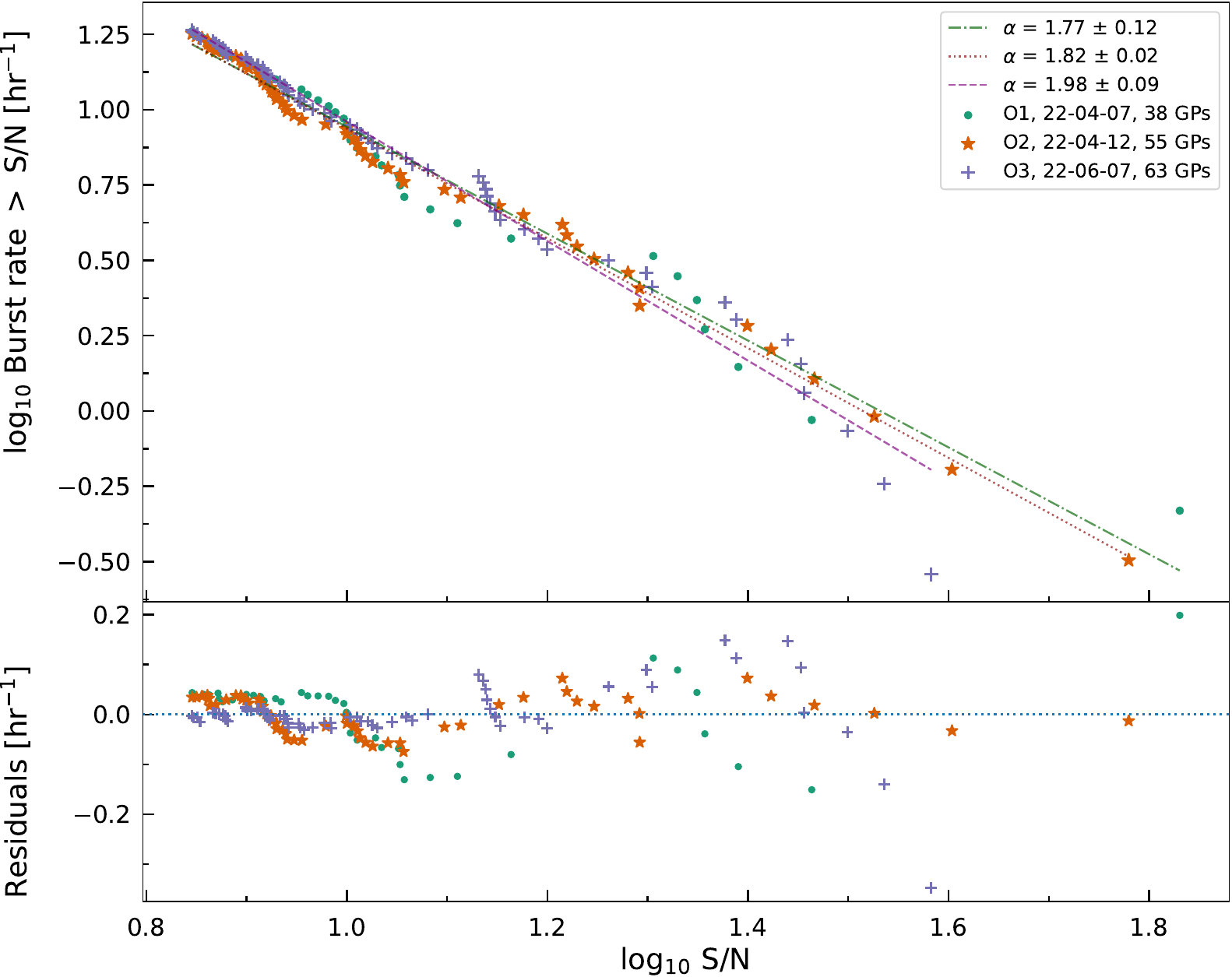}
    \caption{\textit{Upper panel:} Cumulative S/N distributions of the GPs detected in O1 (green dots), O2 (orange stars), and O3 (purple plus-signs). The y-axis shows the number of GPs at a certain S/N, or higher, per hour. Dashed lines represent least-squares best-fit power laws with indices $\alpha = 1.77\pm0.12$ (O1, green, dot-dashed), $\alpha = 1.82\pm0.02$ (O2, orange, dotted), and $\alpha = 1.98\pm0.09$ (O3, purple, dashed). \textit{Lower panel:} Best-fit residuals showing that a power-law fit is a reasonable but not perfect description of the observed distribution. The remaining scatter in the data points is likely due, at least in part, to RFI contamination -- despite masking the frequency channels that are the most corrupted.}
    \label{fig:energy_distr}
\end{figure}

Figure~\ref{fig:energy_distr} shows that the S/N distribution of GPs detected with Parkes follows a steep power law, though with some scatter that may be due to residual RFI in the data. 
The S/N distribution shows no turnover towards low S/N, indicating that our detection method likely captured nearly all GPs above the S/N threshold. Consequently, no additional thresholds have been applied, and the cumulative S/N distributions are fitted with a power law over the entire S/N range using the least-squares method. The best-fit power-law indices are $\alpha = 1.77 \pm 0.12$, $\alpha = 1.82 \pm 0.02$, and $\alpha = 1.98 \pm 0.09$ for O1, O2, and O3, respectively, with uncertainties estimated using bootstrapping.

\subsection{Pulse characteristics}
Table~\hyperref[tab:gpproperties]{2} provides the observed characteristics of the six brightest GPs per observation (an arbitrarily chosen number), labeled, e.g., O1-B01 to O1-B06 (in the case of O1), serving as a sub-sample used for detailed analysis. The times of arrival, S/N values, and other relevant data are presented in the table. The observed GP widths above $2500$\,MHz for the brightest GPs suggest an intrinsic pulse width of $\lesssim 2$\,\mus, consistent with previous studies \citep{knight_2006_apj}. Assuming a consistent width of $2$\,\mus\ for all GPs, we calculated peak flux densities and, subsequently, fluences and isotropic-equivalent spectral luminosities.
The brightest GP, O2-B02, has a fluence of $\mathcal{F} = 49.88 \pm 9.98$\,Jy~\mus, with a corresponding isotropic-equivalent spectral luminosity of $L_{\nu,iso} = (9.63 \pm 1.87) \times 10^{23}$\,erg~s\minusone~Hz\minusone.
   
\begin{table*}
    \caption{Observed properties of the six brightest M28A giant pulses (GPs) per observation, selected as a sub-sample for further analysis. Each GP is labeled with the observation session (O1–O3) followed by a burst number, which resets for each session for clarity (e.g., O1-B01 to O1-B06). The fluence is based on an assumed width of 2 \mus\ for all GPs (radiometer equation, 20\%\ SEFD uncertainty), as the time resolution is insufficient to resolve the GPs at the highest frequencies.}
    \label{tab:gpproperties}
    \small
    \setlength{\tabcolsep}{4pt}
    \begin{threeparttable}
    \centering
        \begin{tabular}{ l c c c c }
            \hline
            \hline
            Giant Pulse ID\tnote{} & Time of Arrival\tnote{a} & S/N\tnote{b} & Fluence\tnote{} & Spectral Luminosity\tnote{c} \\
            & [MJD] & & [Jy~\mus] & [10$^{23}$\,erg~s\textsuperscript{-1}~Hz\textsuperscript{-1}] \\
            \hline
            \multicolumn{5}{l}{O1, 2022-04-06}\\
            O1-B01 & 59676.71818546914 & 22.2 & 14.9 \plm\ 3.0 & 2.8 \plm\ 0.6 \\
            O1-B02 & 59676.75965043257 & 71.3 & 47.8 \plm\ 9.6 & 9.0 \plm\ 1.8 \\
            O1-B03 & 59676.78503077052 & 42.4 & 27.9 \plm\ 5.6 & 5.2 \plm\ 1.1 \\
            O1-B04 & 59676.78662902412 & 19.9 & 13.2 \plm\ 2.7 & 2.5 \plm\ 0.5 \\
            O1-B05 & 59676.79114739647 & 23.1 & 15.4 \plm\ 3.1 & 2.9 \plm\ 0.6 \\
            O1-B06 & 59676.80289653072 & 27.4 & 18.2 \plm\ 3.6 & 3.4 \plm\ 0.7 \\
            \hline 
            \multicolumn{5}{l}{O2, 2022-04-12}\\
            O2-B01 & 59681.74424187666 & 31.6 & 21.4 \plm\ 4.3 & 4.0 \plm\ 0.8 \\
            O2-B02 & 59681.77591744845 & 72.5 & 49.9 \plm\ 10.0 & 9.4 \plm\ 1.9 \\
            O2-B03 & 59681.77869662724 & 31.5 & 21.5 \plm\ 4.3 & 4.0 \plm\ 0.8 \\
            O2-B04 & 59681.78606520491 & 42.7 & 28.9 \plm\ 5.8 & 5.4 \plm\ 1.1 \\
            O2-B05 & 59681.80010849002 & 22.3 & 15.1 \plm\ 3.0 & 2.8 \plm\ 0.6 \\
            O2-B06 & 59681.80305924947 & 51.5 & 35.1 \plm\ 7.0 & 6.6 \plm\ 1.3 \\
            \hline 
            \multicolumn{5}{l}{O3, 2022-06-07}\\    
            O3-B01 & 59737.53309938375 & 40.4 & 27.3 \plm\ 5.5 & 5.1 \plm\ 1.0 \\
            O3-B02 & 59737.55251485373 & 42.1 & 28.7 \plm\ 5.7 & 5.4 \plm\ 1.1 \\
            O3-B03 & 59737.55580092672 & 35.4 & 24.2 \plm\ 4.8 & 4.5 \plm\ 0.9 \\
            O3-B04 & 59737.62478444591 & 38.2 & 25.9 \plm\ 5.2 & 4.9 \plm\ 1.0 \\
            O3-B05 & 59737.63544606505 & 24.4 & 16.6 \plm\ 3.3 & 3.1 \plm\ 0.6 \\
            O3-B06 & 59737.64736762035 & 38.5 & 25.8 \plm\ 5.2 & 4.8 \plm\ 1.0 \\
            \hline
        \end{tabular}
        \begin{tablenotes}
            \item\hspace{4.2mm} \textsuperscript{a} Barycentered at infinite frequency, DM\textsubscript{O1,O2} = 119.914\,\dmunits \&  DM\textsubscript{O3} =
            \item \hspace{5.8mm} 119.906\,\dmunits.
            \item \hspace{4.2mm} \textsuperscript{b} Peak S/N.
            \item \hspace{4.2mm} \textsuperscript{c} Isotropic-equivalent, calculated at a distance of 5.6\,\plm\,0.3 kpc \citep{oliveira_2022_aa}.
        \end{tablenotes}
    \end{threeparttable}
\end{table*}

\subsection{Dynamic spectra}
\begin{figure*}
  \centering
  \begin{minipage}{\textwidth} 
    \includegraphics[width=\textwidth]{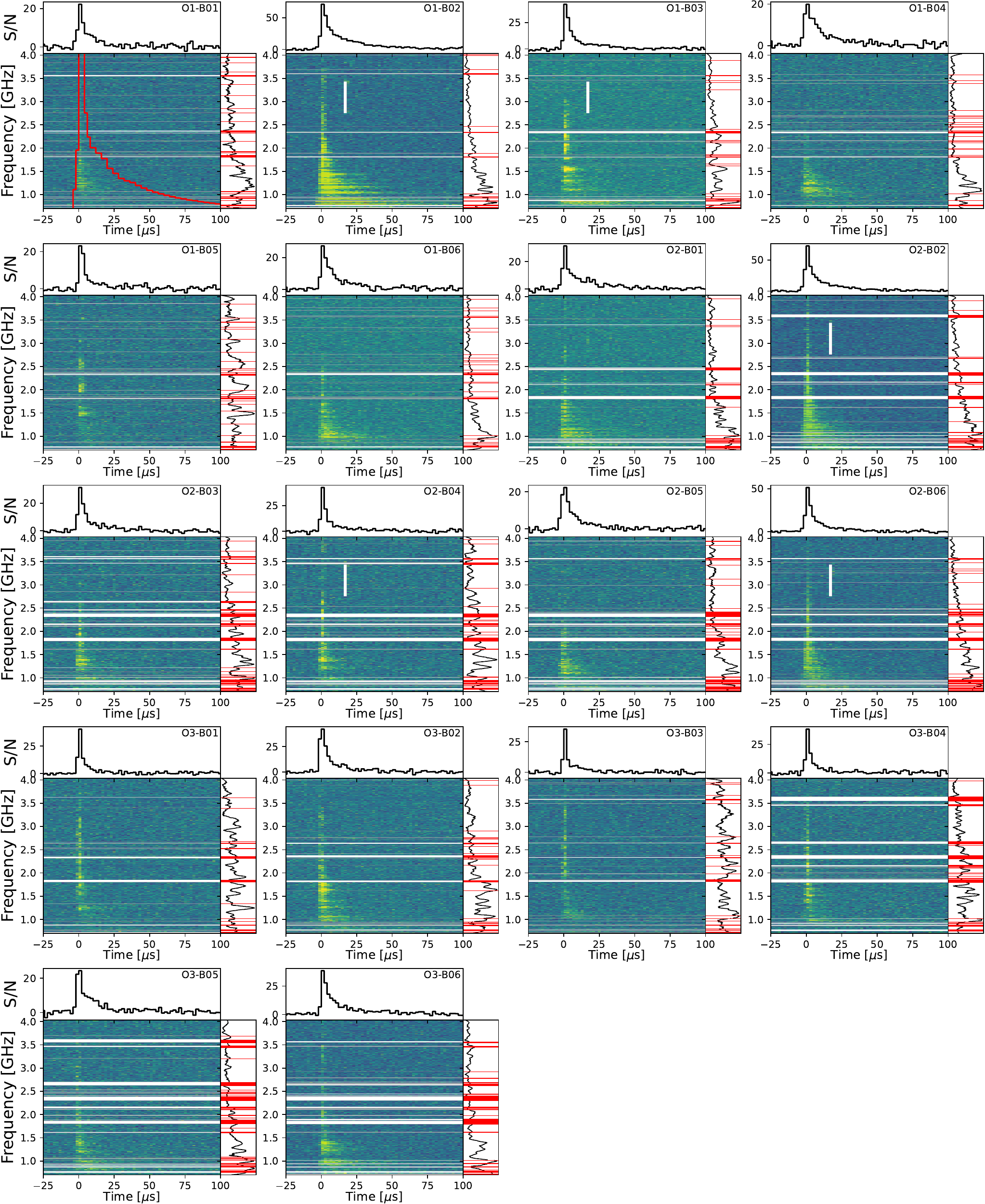}
  \end{minipage}
  \caption{Dynamic spectra displaying the bandpass-corrected Stokes~I values of the six brightest GPs per observation. The frequency/time resolution is $4$\,MHz / $2$\,\mus. Each pulse is dedispersed using a single DM per observation: $\text{DM\textsubscript{O1,O2}}=119.914$\,\dmunits\ and $\text{DM\textsubscript{O3}}=119.906$\,\dmunits. The upper sub-panels show the frequency-averaged profiles. The 1D frequency spectra, shown in the right sub-panels, are obtained using the burst confinement as highlighted in red for O1-B01. A Savitzky-Golay filter with a window size of 16 and polynomial order 2 is applied on these frequency spectra for visual clarity. Red horizontal bars indicate channels that were manually flagged to remove RFI. Vertical white bars in the dynamic spectra are from flagging to conceal echo-like features with a $16$\,\mus\ offset, arising from sporadic instrumental artefacts during O1 and O2 (as documented in the Parkes observation history log). Table\,\ref{tab:gpproperties} presents characteristics of the bright GPs in this sub-sample.}
  \label{fig:family_plot_18GPs_4col}
\end{figure*}

The dynamic spectra of the six brightest GPs per observation are shown in Figure~\ref{fig:family_plot_18GPs_4col}. 
These spectra provide insights into the spectro-temporal characteristics of the GPs. While the GPs generally have steep spectra, occasionally the spectra appear more flat. The GPs consist of multiple bright patches, of which the sizes and frequency of occurrence varies between GPs. In the following sections, we describe the spectra in more detail and quantify observed characteristics.

\subsection{Scattering and scintillation}
The dynamic spectra of the GPs presented in Figure~\ref{fig:family_plot_18GPs_4col} show increasingly broad tails towards lower radio frequencies, reflecting the chromatic nature of scattering. Figure~\ref{fig:scattertail_fit_O1B02} presents the total-intensity profile of the brightest GP, O1-B02, corresponding to a $200$\,MHz sub-band shown for three central frequencies. We obtained least-squares best-fit scattering timescales of $\tau_{\text{scatt}} = 9.51\pm0.61$\,\mus, $\tau_\text{scatt} = 20.77\pm0.99$\,\mus, and $\tau_{\text{scatt}} = 49.24\pm2.26$\,\mus, corresponding to central frequencies $\nu_{\text{cent}} = 1486$\,MHz, $\nu_{\text{cent}} = 1174$\,MHz, and $\nu_{\text{cent}} = 862$\,MHz, respectively.

According to the NE2001 model's `DM to Distance' method\footnote{\url{https://apps.datacentral.org.au/pygedm/}}, using DM$_{\text{M28A}} = 119.914$\,\dmunits, the predicted Milky Way ISM contribution to scattering at $1.5$\,GHz is approximately 8.2\,\mus. Our observations match this prediction well because the NE2001 value is uncertain at the level of at least a factor of a few, and we measured a scattering timescale of \tscatt($1.5$~GHz) = $9.51\pm0.61$\,\mus\ for O1-B02, see Figure~\ref{fig:scattertail_fit_O1B02}.

Figure~\ref{fig:tscat_plfit_O1B02_O2B02_O2B06} shows the scattering timescales as a function of frequency for the three brightest GPs. The best-fit power-law indices, which here corresponds to an approximation of the frequency dependence of scattering, are $\beta = 3.58 \pm 0.29$, $\beta = 1.66 \pm 0.43$, and $\beta = 3.97 \pm 0.41$, for O1-B02, O2-B02, and O2-B06, respectively.

As we measured for O1-B02 a scattering timescale of $\tau_{\text{scatt}}\text{($1174$\,MHz)}=20.77 \pm 0.99$\,\mus, we estimate the expected scintillation bandwidth at $1.2$\,GHz (approximately $1174$\,MHz) to be $\Delta \nu_{\text{scint}}($1.2$\,\text{GHz}) \sim 1/(2\pi\tau_{\text{scatt}}) \sim 8$\,kHz, assuming a single scattering screen.

\subsection{Individual spectra}
\begin{figure}
	\includegraphics[width=\columnwidth]{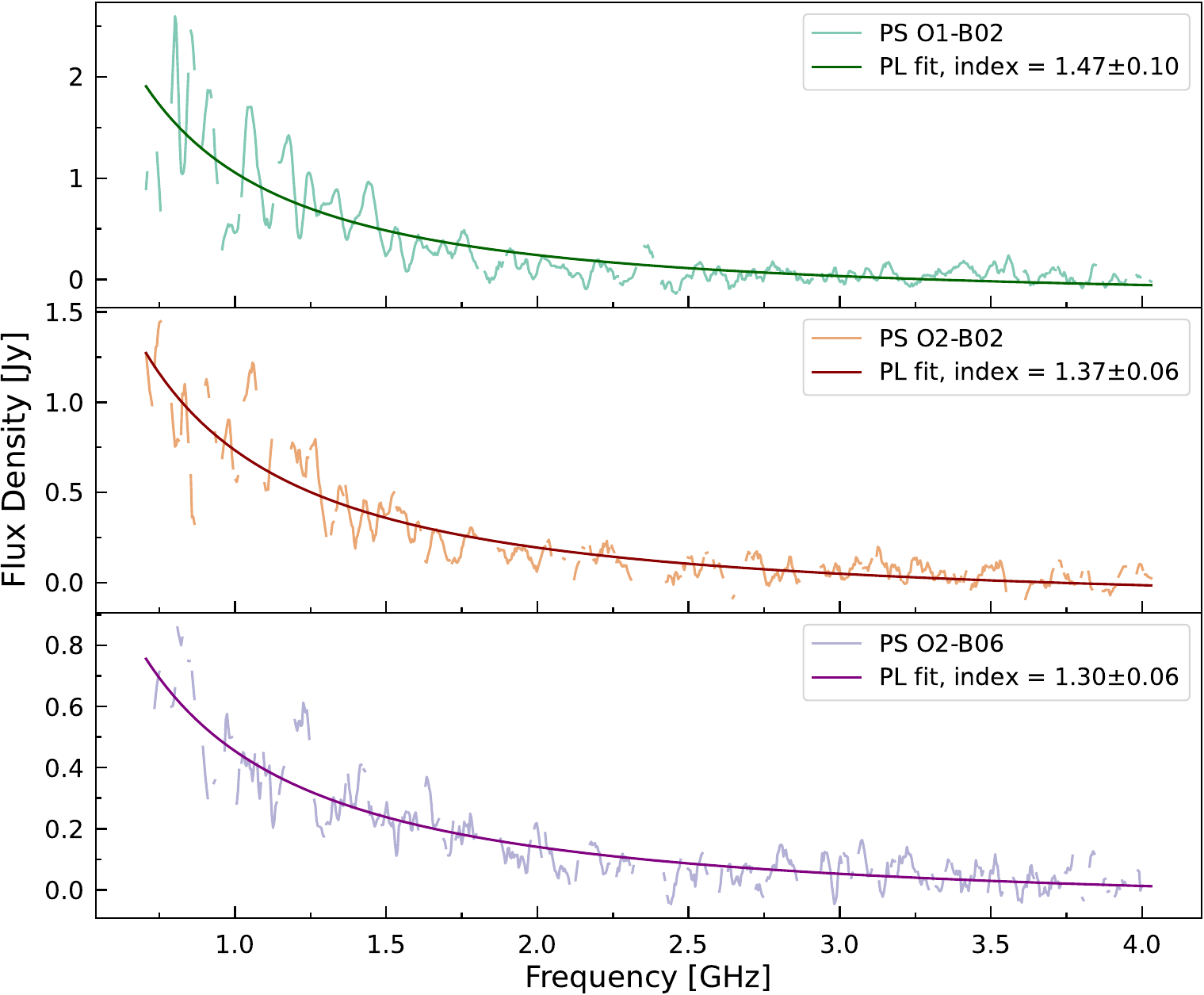}
    \caption{The spectra of the three brightest GPs detected with Parkes, O1-B02 (upper panel, green), O2-B02 (middle panel, orange-red), and O2-B06 (lower panel, purple). Solid lines represent the least-squares best-fit power law, with corresponding indices of $k=1.47 \pm 0.10$, $k=1.37 \pm 0.06$, and $k=1.30 \pm 0.06$, respectively. O1-B02 is brightest at the lower frequencies compared to the other two and shows the steepest spectrum. Due to the patchy frequency structure, variations in S/N, as well as due to RFI-flagging, the steepness of the spectra varies.}
    \label{fig:spectra_fit_O1B02_O2B02_O2B06}
\end{figure}

The spectra generally exhibit increased S/N toward lower frequencies, see, e.g., Figure~\ref{fig:family_plot_18GPs_4col}.
For the three brightest GPs (O1-B02, O2-B02, and O2-B06), and assuming a single power-law of the form $S\propto\nu^{-k}$, we determined least-squares best-fit spectral indices of $k=1.47 \pm 0.10$, $k=1.37 \pm 0.06$, and $k=1.30 \pm 0.06$, respectively, as shown in Figure~\ref{fig:spectra_fit_O1B02_O2B02_O2B06}. Some instances display more flattened spectra, such as O1-B05, O3-B01, or O3-B03 in Figure~\ref{fig:family_plot_18GPs_4col}. Deviations from a simple power-law are likely due to residual RFI and the spectral `bumps' we describe below.

\begin{figure*}
  \centering
  \begin{minipage}{\textwidth} 
    \includegraphics[width=\textwidth]{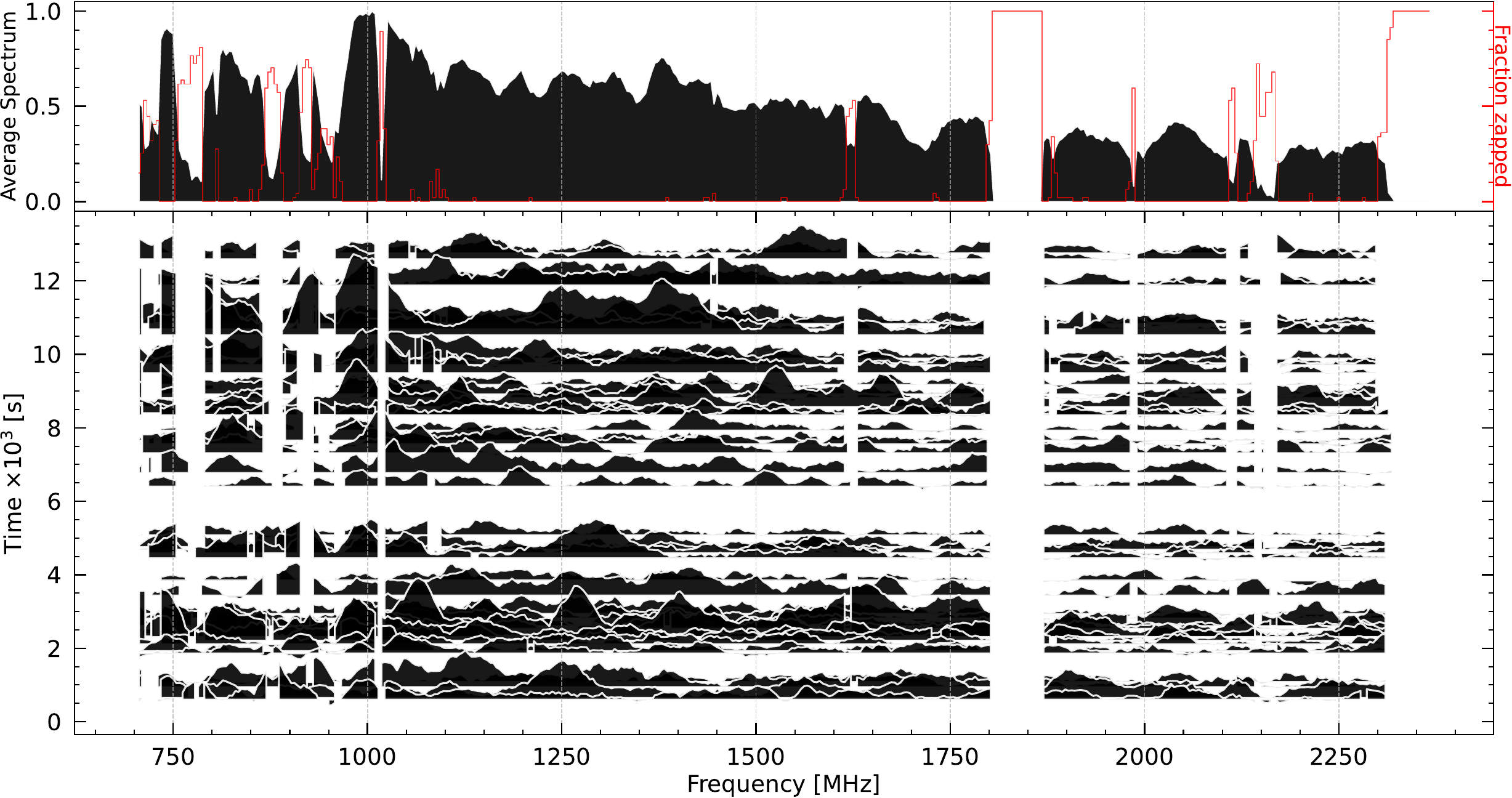}
  \end{minipage}
  \hfill 
  \begin{minipage}{\textwidth} 
    \includegraphics[width=\textwidth]{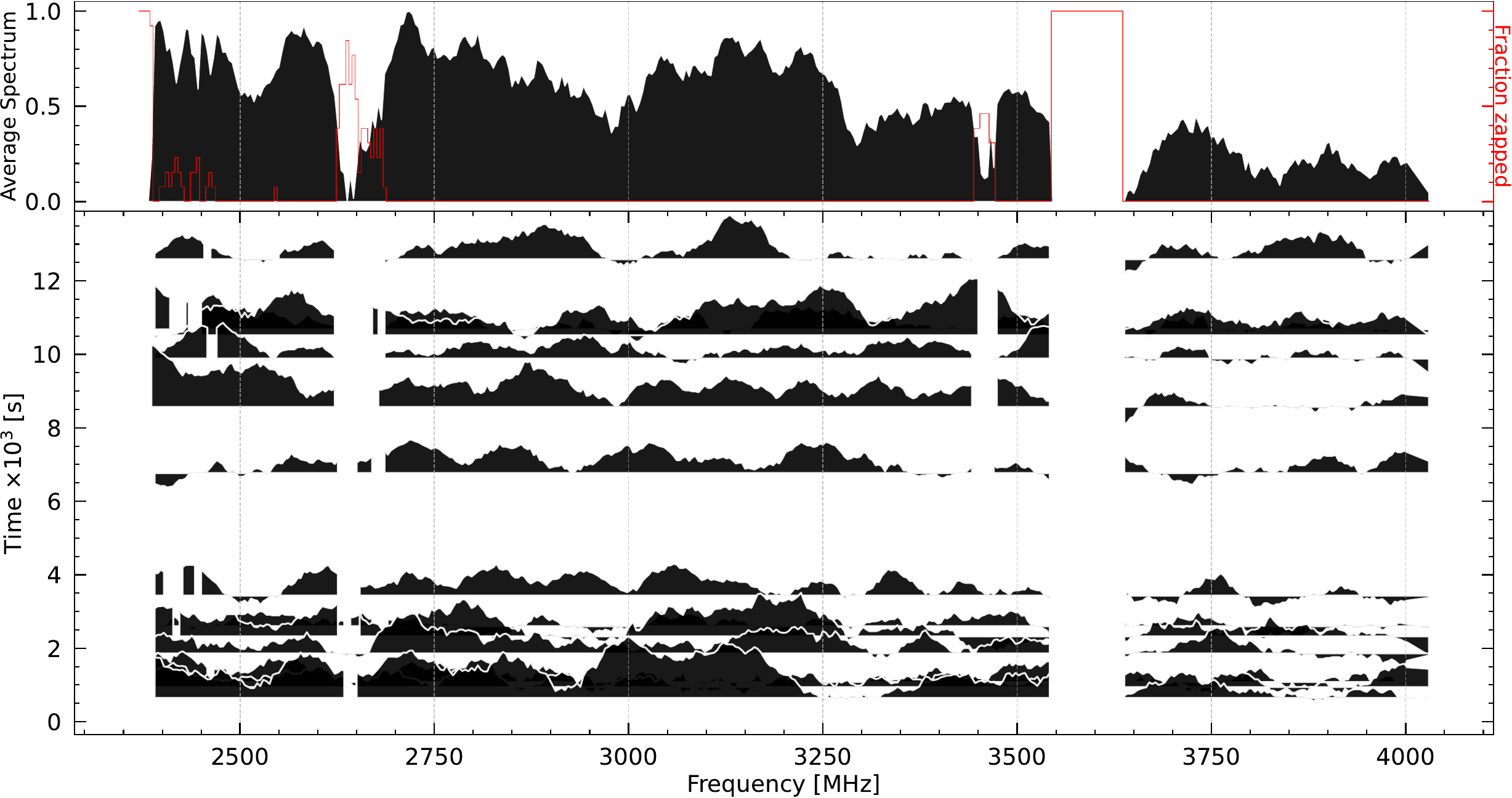}
  \end{minipage}
  \caption{Individual 1D frequency spectra of GPs recorded during O3. A Savitzky-Golay filter has been applied using $wl=8$ and $po = 1$. The top figure displays the spectra within the frequency range of $704-2368$\,MHz, and the bottom figure covers the range from $2368-4032$\,MHz. An additional frequency-averaged S/N threshold of 9.0 is used for both sub-bands, independently. The vertical axis represents the arrival time of each GP relative to the start of the observation, with gaps indicating the absence of GPs above the S/N threshold during certain time intervals. Spectra are amplitude-normalised relative to the brightest GP in the sample. These spectra are then arbitrarily scaled using a single scaling factor for each half of the band, with a larger factor applied to the bottom figure. 
  White vertical stripes denote manually masked regions due to RFI. Each top-panel shows the normalised average spectrum, with the red line indicating the fraction of masked channels due to RFI (across all spectra). For each of the band halves, the average spectrum exhibits increased amplitude towards lower frequencies. The spectra display distinctive patches with sizes on a frequency scale of approximately $10-200$\,MHz. Notably, the GPs do not exhibit preferred frequencies throughout the observation period. In the lower figure, certain spectra exhibit negative amplitudes. This negative amplitude can be attributed to the influence of instrumental noise, despite the data having been corrected for bandpass effects. Figure style inspired by Figure~12 in \citet{hankins_2016_apj}.} 
  \label{fig:JD_spec_obs3}
\end{figure*}

\begin{figure*}
  \centering
  \begin{minipage}{\textwidth} 
    \includegraphics[width=\textwidth]{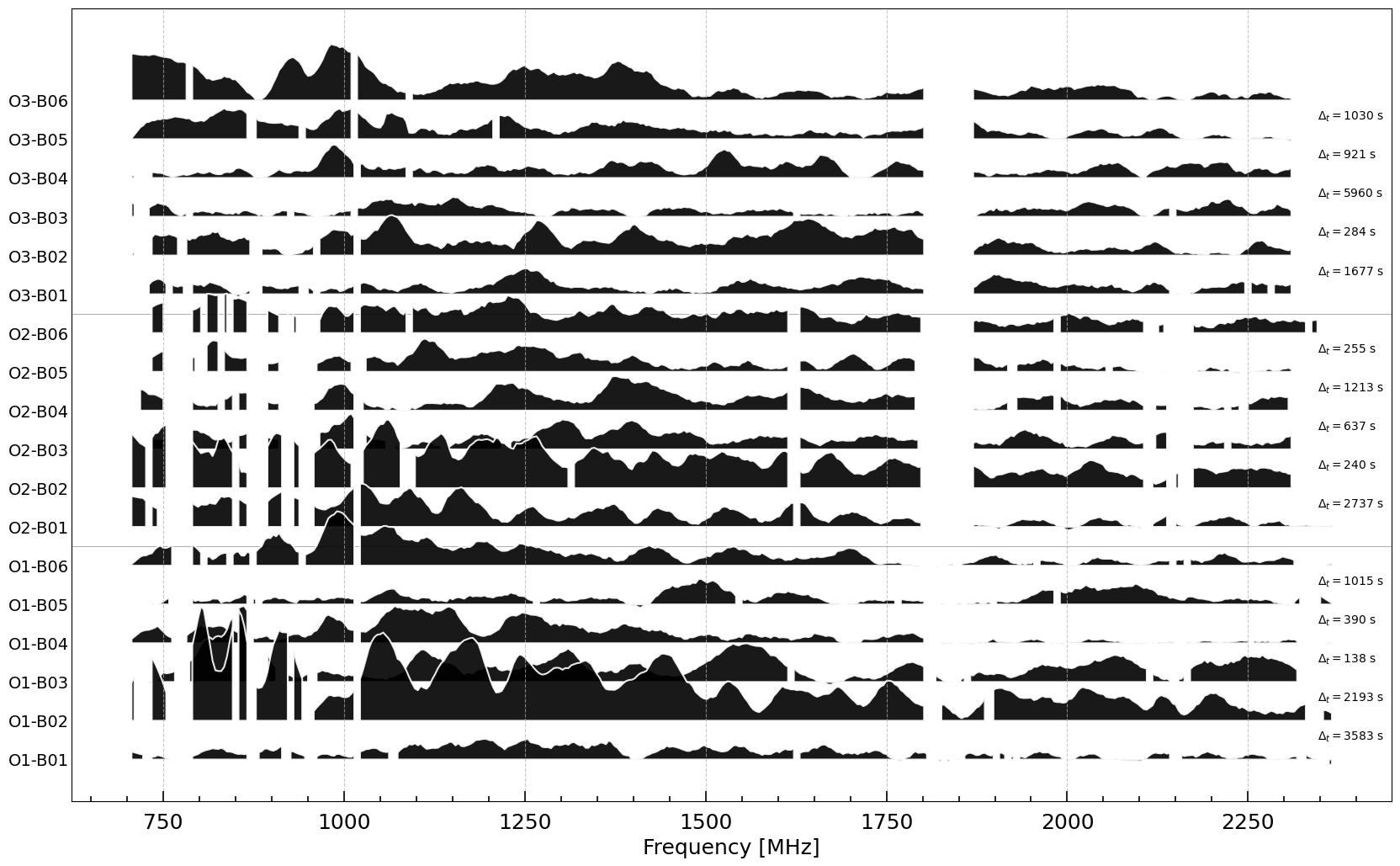}
  \end{minipage}
  \hfill 
  \begin{minipage}{\textwidth} 
    \includegraphics[width=\textwidth]{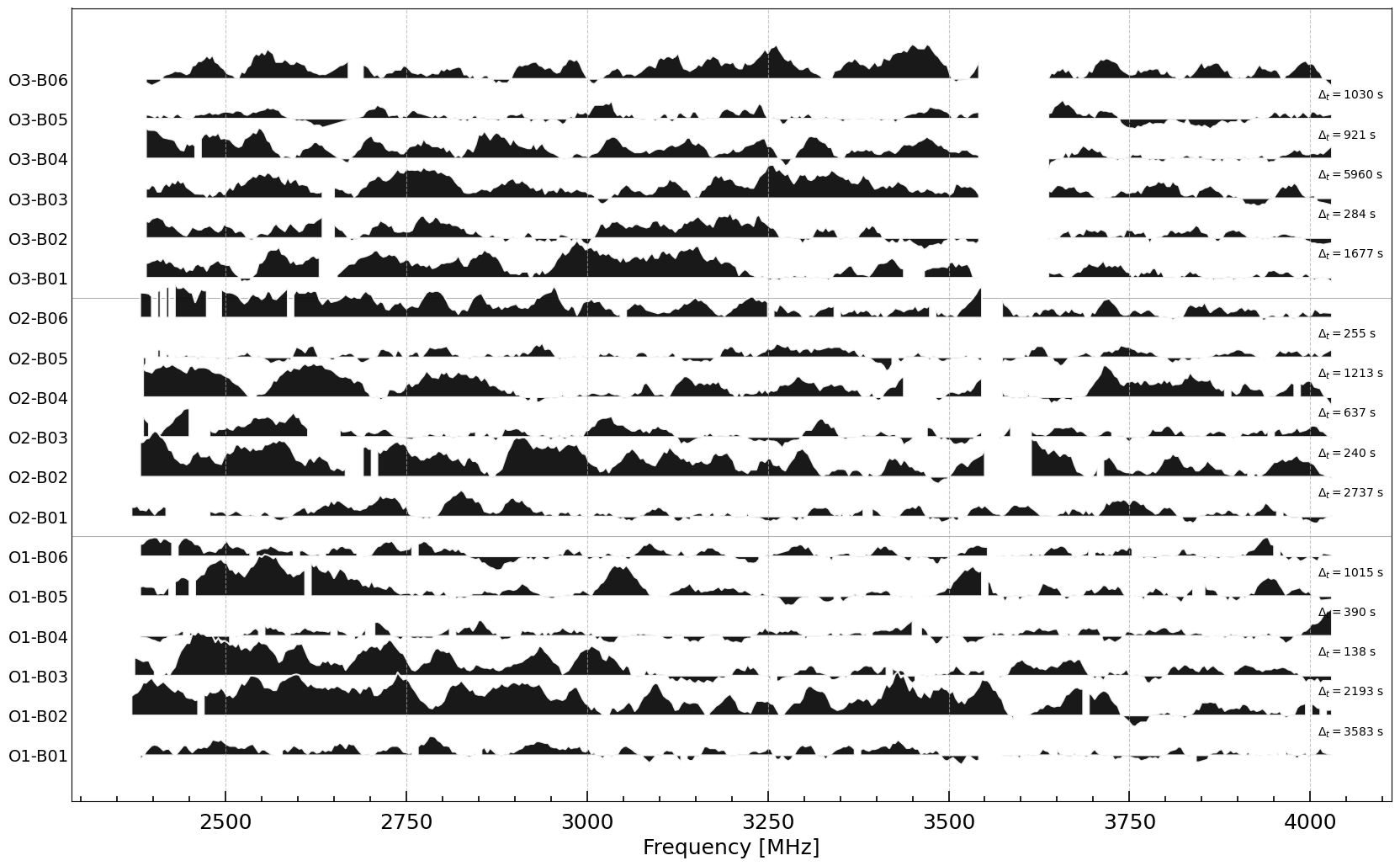}
  \end{minipage}
  \caption{Individual 1D frequency spectra of the six brightest GPs per observation. The spectra are amplitude-normalised to a reference burst O1-B02. These spectra are then arbitrarily scaled, using the same factor for each half of the band, with a larger scaling factor applied to the bottom figure. The Savitzky-Golay filter has been applied ($wl=8$, $po=1$). RFI zapping is denoted by vertical white stripes. The O3 GPs have less strict RFI zapping than shown in Figure~\ref{fig:JD_spec_obs3}. Relative time-of-arrival differences, $\Delta_t$, are indicated on the right side of the figure. Negative amplitudes in the lower figure are due to a too-low S/N at those frequencies and the noise dominating over any astrophysical emission.}
  \label{fig:JD_spec_obs123}
\end{figure*}

Figure~\ref{fig:JD_spec_obs3} shows the individual spectra of M28A GPs detected in O3 as a function of time. We divided the band into two halves ($704-2368$\,MHz and $2368-4032$\,MHz), applying a S/N threshold of $9.0$ for each half separately. All three observations O1 to O3 have similar structures; hence, we present only the GPs from O3 as representative example. Figure~\ref{fig:JD_spec_obs123} shows the spectra of the selection of the six brightest bursts per observation. 

The individual GPs show `bumpy' spectra, with many bright patches across the full bandwidth. The spectral bumps vary in bandwidth, ranging from 10s of MHz to approximately $100-200$\,MHz. We quantified the decorrelation bandwidths by performing a 1D ACF analysis on the spectra of the 18 brightest GPs, visualised in Figure~\ref{fig:acf_O1_O2_O3_B01_to_B06}. The best-fit HWHM values range between $15.49\pm0.27$\,MHz (O1-B02) and $46.73\pm0.29$\,MHz (O1-B03), as listed in the figure.

Sometimes the spectral bumps are blended, and sometimes they appear as isolated narrow-band patches. The locations of these spectral bumps vary in frequency between GPs, with no discernible pattern. While in some instances the spectral bumps appear evenly spaced (see for example O2-B01 or O1-B02 in Figure~\ref{fig:JD_spec_obs123}), we find no strong evidence for a reproducible spacing like the `zebra bands' seen from the Crab pulsar GPs \citep{eilek_2016_jpm, hankins_2016_apj}. 

\subsection{Sub-band search}
\begin{figure*}
	\includegraphics[width=\textwidth]{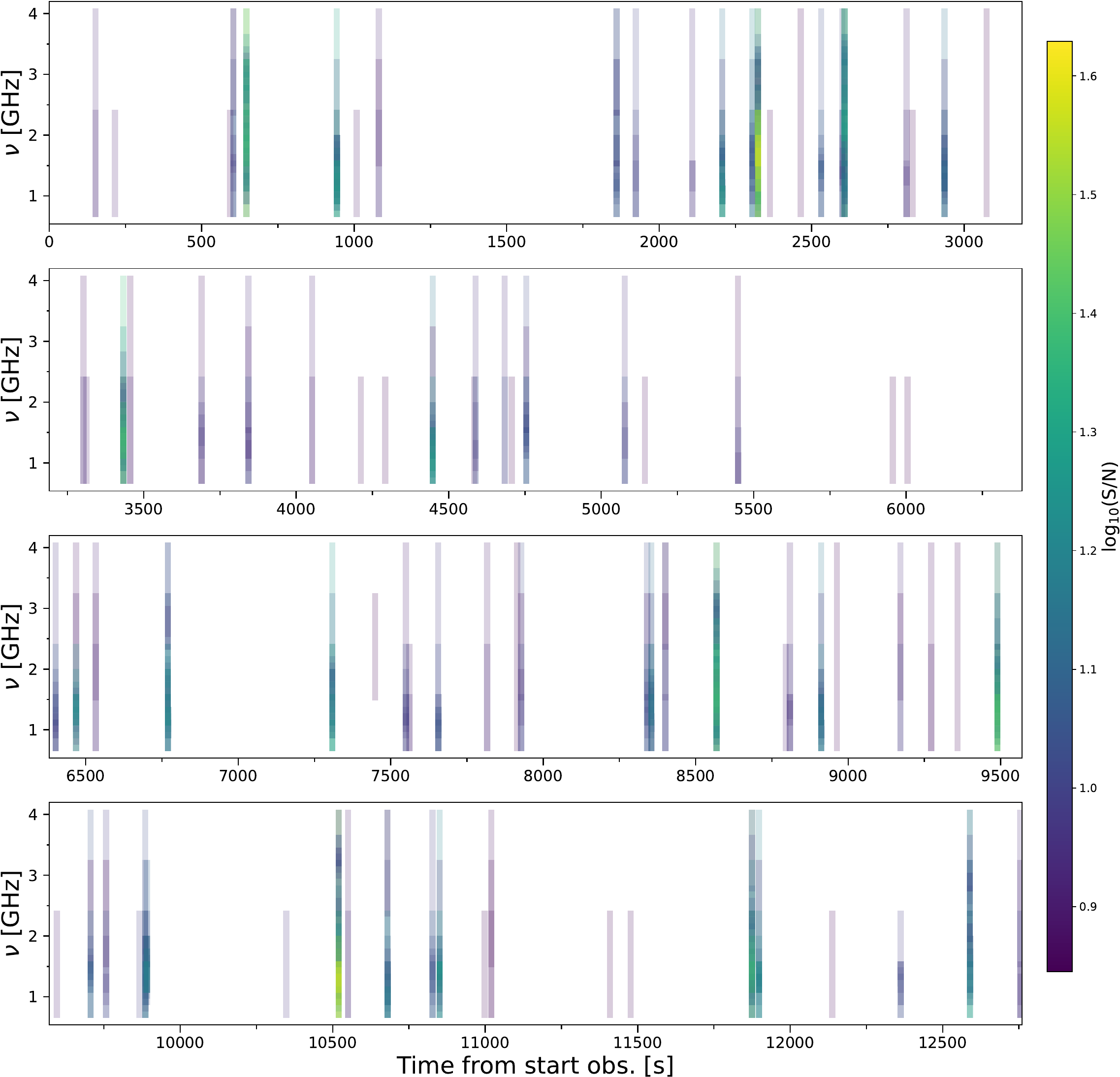}
    \caption{Sub-band search results for O3, showing GP detections in each sub-band as well as in the full-band. GP detections are denoted by vertical rectangles, matching sub-band central frequencies. Colour indicates observed peak S/N (directly measured in each sub-band search) with a logarithmic colour scheme. The x-axis represents time-of-arrival relative to the observation start. Some GPs are within a few minutes of each other, and hence (partially) overlapping with each other in this representation. See Figure~\ref{fig:waittime_distr} for a summary of the wait-time distribution.}% neww
    \label{fig:sbs_results_obs3}
\end{figure*}

Figure~\ref{fig:sbs_results_obs3} shows the outcome of the sub-band search for O3. It shows the sub-band(s) in which a GP was detected, along with its observed peak S/N. Additionally, it indicates full-band GP detections, testing whether the sub-banded search revealed new GPs that might have otherwise been undetected. 

Most GPs are detected in multiple sub-bands and mostly below $2$\,GHz. Sub-band peak S/N appears to be increasing towards lower frequencies. About $20$ GPs are detected in the $1664$-MHz sub-bands that would have been missed in the full-band search. These are, however, very faint having low peak S/N values below $7.5$.

A genuine narrow-band GP ($<200$\,MHz) would show up in Figure~\ref{fig:sbs_results_obs3} as a single sub-band detection, without any broader sub-band detections (or the hypothetical narrow-band GP would have to be very bright to be detectable in broader bands). We do not detect such narrow-band GPs in $3.5$\,hr of data, which we confirmed by examining the dynamic spectrum of each event by eye. Thus, the occurrence rate of strictly narrow-band GPs appears to be low. However, the spectra do show bright patches that could look like narrow-band GPs if the observing band-width would have been smaller. We illustrate this with two examples in the next section. 

For O1 and O2 the $1664$\,MHz sub-band search results are similar to the $1664$\,MHz detection in O3. Therefore, we show the sub-band search results only for O3, as a representative example.

\subsection{Band-limited observations vs broad-band observations}
\begin{figure*}
    \centering
    \begin{minipage}{\textwidth} 
        \includegraphics[width=\textwidth]{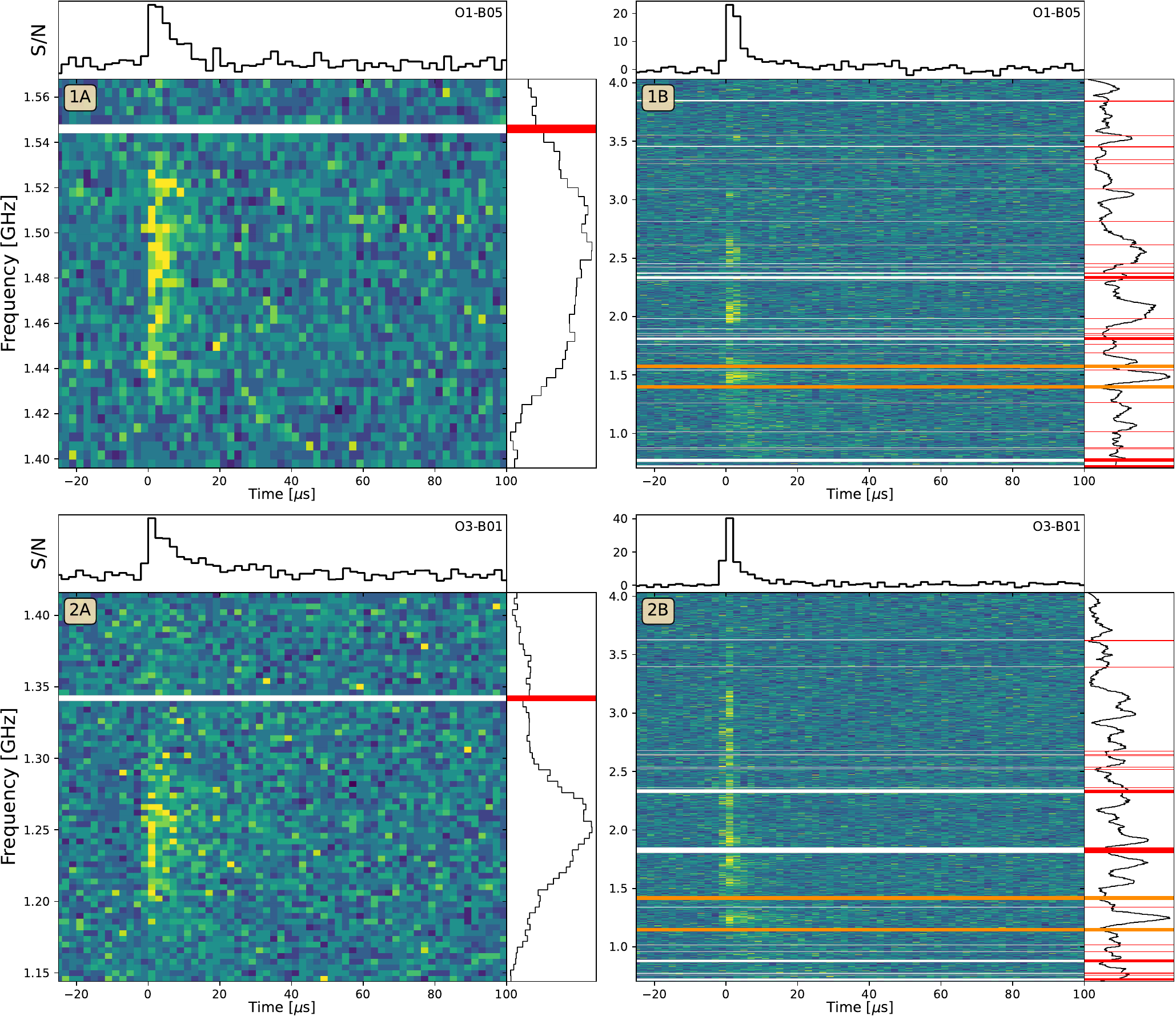}
    \end{minipage}
    \caption{\textit{Top-left (1A)}: Dynamic spectrum of GP O1-B05 with a frequency/time resolution of $4$\,MHz / $2$\,\mus. Missing channels were excised due to RFI, and are marked with a red bar. The top panel displays the corresponding dedispersed time-series, and the right panel shows the spectrum. A prominent narrow-band feature ($\sim 100$\,MHz) is visible between $1.44$ and $1.52$\,GHz. \textit{Top-right (1B)}: The same GP, O1-B05, shown over a larger frequency range, revealing the apparent narrow-band feature as a bright patch within a significantly broader GP profile extending over $2.5$\,GHz. The frequency range of the zoomed-in figure (1A) is highlighted with orange lines. \textit{Bottom-left (2A)}: Dynamic spectrum of GP O3-B01, showing a distinct narrow-band feature. \textit{Bottom-right (2B)}: The same GP, O3-B01, shown over a larger frequency range, demonstrating that the apparent narrow-band feature is in fact a bright patch within a much broader GP profile spanning over $2.5$\,GHz.}
    \label{fig:dynspec_O1B05_O3B01_zoominzoomout}
\end{figure*}

Figure~\ref{fig:dynspec_O1B05_O3B01_zoominzoomout} shows two examples of narrow-band ($\sim200$\,MHz) patches. It shows the dynamic spectrum zoomed-in around the patches, mimicking a band-limited observation in which the narrow-band patch appears to be an isolated narrow-band GP. The figure shows the zoomed-out dynamic spectra as well, illustrating that the wide-band observations reveal that the narrow-band patches are in fact part of a larger GP structure.

\subsection{DM variability}
Figure~\ref{fig:snrdm123} displays the DM-S/N density distribution for the three Parkes observations, using optimised S/N and corresponding DM values determined by \verb|PRESTO|'s \verb|single_pulse_search.py|.
We calculated best-fit DM values of $119.915$\,\dmunits, $119.914$\,\dmunits, and $119.906$\,\dmunits\ for O1, O2, and O3, respectively, using \pam| on an archive file of 10 minutes of folded data. These values closely match the average densities shown in Figure~\ref{fig:snrdm123}.

When comparing the DM values of O1 and O2, which are five days apart, we observe an apparent DM variation of approximately $0.001$\,\dmunits. Similarly, when comparing O1 and O2 with O3 (which is two months apart from O1 and O2), we find a DM decrease on the order of $0.01$\,\dmunits.

Figure~\ref{fig:dm_longterm_var} presents the \pam|-fitted DM values for the Parkes GP sample versus the date of observation. The figure also includes data points from our previous study of M28A's GPs, which we conducted using archival Green Bank Telescope (GBT) data \citep{vanruiten_2023_mscthesis}. To provide context on longer timescales, we also include DM determinations from one to three decades ago, as reported by \citet{cognard_1997_aa}. Over a 30-year period, the DM increased by approximately $\sim 0.1$\,\dmunits, suggesting a crude estimate of an average long-term DM increase of $\Delta \text{DM}/dt \sim 0.003$\,\dmunits~yr\minusone.

\section{Discussion} 
\label{sec:discussion}

Here we discuss our results in the context of previous studies of M28A \citep{knight_2006_apj,bilous_2015_apj} and compare with the repeating \frb.

\subsection{Comparison with previous M28A studies}
\label{sec:M28Astudies}

\subsubsection{Durations}
In our analysis, we used the native recorded time resolution of $2$\,\mus\ for all individual GPs. At frequencies above $2200$\,MHz, where the scattering timescale is $\lesssim 2$\,\mus, see Figure~\ref{fig:tscat_plfit_O1B02_O2B02_O2B06}, we observe narrow GP widths of only $1-3$ time bins. Consequently, the GPs are mostly unresolved by our data at high radio frequencies, indicating intrinsic durations of $\lesssim 2$\,\mus, and consistent with previous findings in the literature \citep{knight_2006_apj}. 

\citet{knight_2006_apj} observed GP profiles binned at $750$\,ns, measuring FWHMs (at $1.5$\,GHz) between $750$\,ns and $6.0$\,\mus. Some GPs were unresolved even with a $750$-ns time-resolution. Figure~4 in their paper shows the presence of micro/nano-structure in two GPs, using a time-resolution of $7.8125$\,ns at $2.7$ and $3.5$\,GHz, as well as a pulse at $3.5$\,GHz with a FWHM of $20$\,ns, appearing to be resolved as a single coherent pulse (which could be interpreted as a nanoshot). These observations suggest that at least a fraction of the GPs observed in our dataset are unresolved. For consistency in fluence and isotropic-equivalent spectral luminosity calculations, we assume an equal intrinsic pulse width of $W = 2$\,\mus\ for all GPs. 

\citet{knight_2006_apj} observed a pulse at $3.5$\,GHz being broader than at $2.7$\,GHz, and argue this could indicate resolution of the pulse envelopes. Since this broadening does not scale with frequency, it may reflect variation in the number of nanoshots across the band. The possible role of (interfering) nanoshots in shaping the observed GP spectra is discussed further in \S\ref{sec:spec_analysis}.

\subsubsection{Luminosities}
The brightest GPs in our sample have peak flux densities on the order of tens of Jy. Assuming an equal intrinsic pulse width of $2$\,\mus\ for all GPs, we infer isotropic-equivalent spectral luminosities on the order of $10^{23}$\,erg\minusone~s\minusone~Hz\minusone. 
The actual peak luminosity could be much higher, since we are likely not resolving the GP envelopes and therefore measure smeared-out peak flux density. This makes direct comparisons with previous peak brightness determinations challenging. 
\citet{knight_2006_apj}, using a $64$-MHz observing bandwidth at $3.5$\,GHz with a $250\times$ shorter sampling time ($7.8125$\,ns), observed a GP appearing to be resolved into a single coherent shot with an FWHM of $20$\,ns. For this shot, they measured a peak flux density of $2300$\,Jy, which would corresponds to an isotropic-equivalent spectral luminosity on the order of $10^{25}$\,erg\minusone~s\minusone~Hz\minusone (two orders of magnitude greater than our observations). For this GP-nanoshot, \citet{knight_2006_apj} inferred a brightness temperature $T_\text{B}$ on the order of $10^{37}$\,K.

For GP O1-B02, we measured a peak flux density of $24$\,Jy. Assuming a pulse width of $2$\,\mus, the corresponding brightness temperature (following Equation~13 in \citet{petroff_2019_aarv}) is on the order of $10^{32}$\,K. If we adopt a pulse width of $20$\,ns, similar to what \citet{knight_2006_apj} observed for one of their GPs, $T_\text{B}$ becomes approximately $10^{36}$\,K, a comparable value. The actual $T_\text{B}$ might be a few orders of magnitude higher, considering that the observed peak flux density with a 2\,\mus\ sampling time could potentially be significantly higher with a nanosecond-scale sampling time.

\subsubsection{S/N distribution}
M28A's GPs follow a steep power-law S/N distribution. If the pulse width and effective bandwidth are approximately constant between GPs, the burst energy scales linearly with the peak flux density and therefore with S/N. Under this assumption, the power-law index of the S/N distribution is identical to that of the energy distribution.
The best-fit power-law indices are $\alpha = 1.77\pm 0.20$, $\alpha = 1.82 \pm 0.06$, and $\alpha = 1.98 \pm 0.20$ for O1, O2, and O3, respectively, as illustrated in Figure~\ref{fig:energy_distr}. These results are consistent with the energy-distribution power-law index of $\alpha = 1.8 \pm 0.3$ reported by \citet{bilous_2015_apj} and comparable to the value of $\alpha = 1.6$ determined by \citet{knight_2006_apj}.

In the context of radio transient energy distributions, it is common to observe a flattening at the low-energy end, often attributed to observation sensitivity and methodological limitations \citep[see, e.g.,][]{kirsten_2024_natas}. Such distributions only truly reflect the behaviour of the source after the turn-over point, when the distribution becomes a power law. 

Both \citet{knight_2006_apj} and \citet{bilous_2015_apj} observe this turnover in their data. Notably, our GP sample shows no evidence for such a turnover, and shows no indication of incompleteness down to the S/N threshold. A maximum-likelihood method of the power-law slope $\alpha$ \citep{crawford_1970_apj, james_2019_mnras} indicates that the distribution is consistent with a single power law above this threshold, see Figure~\ref{fig:waittime_distr}. The absence of flattening in our case suggests that the GPs are well-represented even close to the S/N threshold. This could be because there are very few false-positives due to RFI on the few-bin timescale. The difference compared to \citet{knight_2006_apj} and \citet{bilous_2015_apj} may reflect differences in detection thresholds and selection criteria. This difference, however, does not affect the measured PL index.

We do not find any flattening of the distribution towards higher energies, which has been observed for another GP-emitting source, the first-discovered MSP, PSR~B1937+21 \citep{mckee_2019_mnras}. \citet{mckee_2019_mnras} observed flattening in the distribution of their main-phase GPs after combining multiple observations. 
% a flattening in their main-phase GPs after combining multiple observations. 
However, upon combining our data from all three observations, we do not find evidence of flattening towards higher energies for M28A's GPs. 

\subsubsection{Wait-time distribution}
The wait times of the GPs follow a Poisson distribution (see Figure~\ref{fig:waittime_distr}), consistent with previous findings by \citet{knight_2006_apj} and \citet{bilous_2015_apj}. This reflects that the pulses occur randomly and independently in time. We find that the average GP rate as well as the mean wait time does not vary significantly between observations (O1 and O2 are separated by a week, and O2 and O3 by two months). Comparing mean wait times and GP rates with other studies is not trivial, since those quantities depend on both telescope sensitivity, observing band, and observation duration.

\subsubsection{DM variation}
\citet{cognard_1997_aa} analysed DM variations towards M28A using measurements spanning three years from \citet{backer_1993_apj} and their own measurements collected over five years. They observed a linear increase in DM over a six-year period, with secular variations reaching up to $0.005$\,\dmunits~yr\minusone. Our analysis indicates a continued linear increase in the average DM on timescales of decades, estimating a 30-year averaged DM increase of roughly $\Delta \text{DM}/dt \sim 0.003$\,\dmunits~yr\minusone. This long-term DM variation can be explained by the proper motion (PM) of the MSP and the M28 globular cluster, causing our line-of-sight through the ISM to change \citep{cognard_1997_aa}. For M28A, $\text{PM}_{\text{RA}}=-0.28$\,mas/yr and $\text{PM}_{\text{DEC}}=-8.92$\,mas/yr \citep{vasiliev_2021_mnras}, corresponding to a transverse velocity of $ v_{\text{tr}}= 227.21$\,km/s, as listed in the ATNF Pulsar Catalogue\footnote{\url{https://www.atnf.csiro.au/research/pulsar/psrcat/}} \citep{manchester_2005_aj}.

Figure~\ref{fig:snrdm123} shows the DM-S/N density distribution for the three Parkes observations, suggesting a potential form of DM bi-modality at each epoch ($119.914$~\dmunits\ and $119.916$~\dmunits\ for O1/O2, and $119.906$~\dmunits\ and $119.908$~\dmunits\ for O3). This could be due to the fact that bursts that are brighter at lower frequencies are more scattered, and the single-pulse search then compensates with a slightly higher DM to sweep the scattering tail back under the burst to make it narrower and maximize band-integrated S/N.

Alternatively, this could be attributed to the GPs arriving in two distinct phase windows. Speculatively, the GPs may have different frequency extent and spectral indices based on phase of arrival, resulting in a slightly different DM. Splitting the total GP sample in both phase samples is needed to examine this. 

\subsubsection{Spectral analysis}
\label{sec:spec_analysis}
The sub-banded single-pulse search did not detect isolated, \textit{genuinely} narrow-band ($\lesssim 200$\,MHz) GPs (Figure~\ref{fig:sbs_results_obs3}). However, occasionally, the GPs consist of narrow-band `islands' that might look like narrow-band GPs for smaller observing bandwidths (Figure~\ref{fig:dynspec_O1B05_O3B01_zoominzoomout}). Indeed, these islands appear similar to the few instances of apparently narrow-band GPs observed from other pulsars \citep{geyer_2021_mnras, thulasiram_2021_mnras, bij_2021_apj}. 

Most GPs in our sample appear to be broad-band and, notably, we show that the brightest GPs can span the full frequency range from $704$ to $4032$\,MHz (Figure~\ref{fig:family_plot_18GPs_4col}). Furthermore, we find that the decorrelation bandwidth of the Gaussian-peaked islands of $10-100$\,MHz, previously reported by \citet{bilous_2015_apj} for an $800$-MHz bandwidth, persist at a bandwidth of $3.3$\,GHz. \citet{bilous_2015_apj} found that for some sufficiently bright GPs in their $1.1-1.9$\,GHz observations the size of the islands did, however, not exceed $50$\,MHz, and mentioned that a larger number of pulses is needed to confirm any statistical trend. We show that some GPs do have larger sizes at higher frequencies. See, for example, O2-B04 in Figure~\ref{fig:JD_spec_obs123}, having islands with sizes of about $150$\,MHz around $2200-2800$\,MHz. ACFs of the upper- and lower-half of the band suggest that spectral islands become broader towards lower frequencies. However, overlapping patches and RFI-zapping at the lower frequencies complicate any robust claims. 

Our observations also confirm that M28A's GPs show significant spectral variability between pulses \citep{bilous_2015_apj}. From pulse to pulse, the decorrelation bandwidth of the islands varies, and the peaks of the islands do not have preferred frequencies (Figure~\ref{fig:JD_spec_obs3}). Within a single GP spectrum, islands mostly appear with random spacing. Occasionally, the islands display an apparently evenly spaced pattern -- see, e.g., O2-B01, O1-B02 or O1-B05 in Figure~\ref{fig:JD_spec_obs123}. This could well be coincidence, however, given that the spectral peaks appear to be randomly spaced in most cases.

Averaging over the spectral islands, the three brightest GPs have steep power-law spectra (Figure~\ref{fig:spectra_fit_O1B02_O2B02_O2B06}). Upon examination by eye, most of the 18 brightest GPs (Figure~\ref{fig:family_plot_18GPs_4col}) show steep power-law spectra, with occasionally flatter spectra observed (e.g., bursts 01-B05 and 03-B01). Previous studies of M28A used more limited observing bandwidths, complicating single GP spectral index estimates. Therefore, our spectral index estimates of the brightest GPs cannot be compared with earlier work. 
The observation that most GPs have a steep spectrum while some show a flatter spectrum could be explained by the suggestion of \citet{hankins_2003_natur} and \citet{hankins_2007_apj} that GPs could consist of multiple nanoshots that vary in both frequency extent and central frequency. This variability could cause some GPs to appear flat (not extending to the lowest frequencies), while others appear steep (extending to the lowest frequencies). See also discussion (\S4.5) in \citet{bilous_2015_apj}.

For O1-B02, O2-B02, and O2-B06 we determined the frequency dependence of scattering, expressed as the beta-parameter ($\tau_\text{scatt}\propto \nu^{-\beta}$). The estimated scattering parameter for O2-B06 ($\beta = 3.97 \pm 0.41$) matches the predicted $\beta \sim 4.0 - 4.4$ for the simple model of a single, infinitely wide, thin scattering screen and Kolmogorov scattering \citep{bhat_2004_apj}. O1-B02 has a slightly lower estimated value ($\beta = 3.58 \pm 0.29$), and O2-B06 apparently much lower ($\beta = 1.66 \pm 0.43$). 

The low value of $1.66$ might be due to a combination of variation in spectral structure (some GPs are less bright at the lower frequencies) and RFI zapping (which is most prominent at the lower frequencies). The observed widths at the lower frequencies strongly determine the shape of the curve of $\tau_\text{scatt}$ as a function of frequency and thus the resulting best-fit power-law index. The implication is that this complicates comparison with the predicted $\beta \sim 4.0 - 4.4$ in the single-screen scenario.

We measured a scattering timescale of $\tau_{\text{scatt}}\text{(1.2 GHz)}=20.77 \pm 0.99$\,\mus\ for O1-B02, one of our highest-S/N bursts with a clear scattering tail, and assume it to be generally representative of the scattering timescale of M28A. Consequently, based on the requirements for scintillation to occur in the case of single screen, with $\Delta \nu_{\text{scint}} = 1/(2\pi \tau_{\text{scatt}})$, we would expect a scintillation bandwidth of approximately 8\,kHz. For this reason, the observed decorrelation bandwidth ($10-100$\,MHz) is most likely not due to scintillation for the single screen scenario. Additionally, the  frequency resolution of $4$\,MHz makes us insensitive to any structure smaller than $4$\,MHz.

Could the observed peaked spectral structure nonetheless be due to scintillation, in the case of a more complex scenario with for example a finite screen or multiple screens? We suggest that this is unlikely, for mainly two reasons. Firstly, scintillation variations usually occur on timescales much longer than what we observe. We find no connection or link in structure from pulse to pulse, even when separated by only minutes (Figure~\ref{fig:JD_spec_obs123}). Secondly, ACFs of the spectra in the upper- and lower-half of the band hint at a slight increase in decorrelation bandwidth towards lower frequencies, which is the opposite to what happens for scintillation, where the bandwidths increase towards higher frequencies.

Furthermore, the GPs are emitted in phase P1 and P2 \citep[as defined in Figure~\ref{fig:average_profile_M28A};][]{knight_2006_apj, bilous_2015_apj}. While the average profile shows a broader frequency extent in P2 than in P1, \citet{bilous_2015_apj} found no compelling phase-dependent differences in GP properties. Therefore, we did not conduct a phase-resolved analysis here, though such a study could be worthwhile in future work.

In summary, our broad-band observations reveal four significant insights into the spectra of M28A, expanding upon previous studies:
\begin{enumerate}
    \item M28A does not emit genuinely narrow-band GPs, although $100$\,MHz-wide `islands' are observed occasionally, that could be interpreted as `narrow-band' GPs when using limited observing bandwidths.
    \item In general, the higher the observed peak S/N, the broader the observed frequency extent, with the brightest GPs spanning the full $3.3$\,GHz range. While those GPs have steep spectra, some GPs have flatter spectra with lower S/N at the lowest frequencies compared to higher frequencies.
    \item The spectra of the GPs show distinct and overlapping bumps with sizes of $10-200$\,MHz, which vary in their central frequencies from pulse to pulse. No preferred frequencies are generally observed, with the entire $3.3$\,GHz observing band being covered. Within individual GPs, these bumps typically appear with random spacing. Occasionally, spectral peaks appear evenly spaced over the band, although this could be a coincidence given that most pulses show no such pattern. 
    \item The characteristic frequency scale (ACF-determined) appears to be slightly increasing towards lower frequencies.
\end{enumerate}
 
GP banded emission, often characterized by overlapping Gaussian-peaked substructures, is also observed in other sources such as Crab GPs (see \citet{hankins_2016_apj} or Figure 1 in \citet{bij_2021_apj}). A few GP-emitters, such as \psrgeyer, occasionally reveal clear isolated patches with similar scales as M28A's GP substructure patches ($\sim10-200$\,MHz), as shown in Figure~5 of \citet{geyer_2021_mnras}. Similar apparent isolated patches have also been observed for the Crab \citep{thulasiram_2021_mnras}. Some patches observed in M28A are similar to those seen in Crab GPs, supporting the notion that GPs may consist of several unresolved, narrow-band nanoshots with a notable degree of polarization, as discussed by \citet{bilous_2015_apj}. However, isolated narrow-band GPs appear to be rare (which could be due to the limited sample size of known GP sources), and such features and their underlying causes remain subject to further investigation.

\citet{knight_2006_apj} showed that M28A's GPs can consist of multiple nanoshots, as unresolved spikes ($<7.8$\,ns), some merging together forming a $200$-ns burst, similar to what \citet{hankins_2007_apj} observed for Crab main GPs. \citet{bilous_2015_apj} favour the model of nanoshots explaining the narrow-band emission observed for M28A's GPs. They suggest that isolated and overlapping narrow-band patches in M28A's GPs could be attributed to the interference of nanoshots, similar to Crab main GPs and the presence of discrete, unresolved, narrow-band nanoshots. \citet{bij_2021_apj} also discuss the possibility that narrow-band features in Crab GPs could arise from the interference of nanoshots, nearly equally spaced. They discuss that these nanoshots may be echoes produced by plasma lensing in the pulsar wind. However, distinguishing between a relativistically moving blob emitting a banded spectrum and multiple nanoshots is challenging and perhaps even impossible.

\subsection{Comparison with \frb\ and other FRBs}
\label{sec:comparison_M81R_and_other_FRBs}
\citet{cordes_2016_mnras} and \citet{connor_2016_mnras} proposed that particularly bright GPs, termed `super' GPs, emitted by extragalactic pulsars might explain some fraction of the observable FRBs, particularly those in the nearby Universe. While the exact emission mechanisms of GPs and FRBs remain uncertain, most GP models differ from the often-invoked magnetar model for FRBs. The primary distinction is that GPs originate from rotationally powered neutron stars, whereas the energy budget of magnetars is by definition dominated by magnetic energy release. However, certain GP models, such as magnetic reconnection beyond the light cylinder \citep{philippov_2019_apjl}, suggest a possible similarity in emission mechanisms between GPs and FRBs, independent of primary power source.

Although some observational links between pulsars and FRBs exist, the gap in luminosity between pulsars and FRBs has not yet been bridged, see Figure~3 in \citet{nimmo_2022_natas}. At the top of this gap are bursts from \frb, currently the closest-known FRB, at $3.6$\,Mpc \citep{bhardwaj_2021_apjl}. \frb\ stands out as an outlier among active repeaters due to its unusually short burst durations ($50-100$\,\mus) -- with temporal structures on the order of $ \sim 100$\,ns and $\sim 1$\,\mus\ -- and significantly lower isotropic-equivalent spectral luminosities ($100\times$ lower). The localisation of \frb\ to a globular cluster \citep{kirsten_2022_natur} could suggest that the source is an energetic MSP because such sources are formed at a high rate, per unit stellar mass, in such environments. M28A is the highest-energy MSP known in the Milky Way, which is why we compare it's properties \citep[e.g.,][]{nimmo_2022_natas,nimmo_2023_mnras} to \frb, see Table~\ref{tab:comparison_m28a_m81r}.

\frb\ bursts observed by \citet{nimmo_2023_mnras} have isotropic-equivalent spectral luminosities on the order of $10^{28}$\,erg~s\minusone~Hz\minusone. Using the cumulative S/N distribution we observed for GPs from M28A \citep[$d_{\text{M28A}} = 5.6$\,kpc;][]{oliveira_2022_aa}, we estimate a wait time of $\sim 3.0$\,Myr to detect a GP from M28A with an \frb-like luminosity. This estimation is based on simple power-law extrapolation and assumes a consistent, single power-law index, an energy-independent burst duration and bandwidth, and no high-energy cutoff prior to the GPs reaching \frb-like luminosities. If all assumptions hold, then the chances of detecting extragalactic GPs from an M28A-like pulsar are low. 

However, some GP-emitters show a broken power-law GP energy distribution. \citet{mckee_2019_mnras} analysed GPs from PSR~B1937+21, and found a broken power-law cumulative energy distribution \citep[see Figure~6 in ][]{mckee_2019_mnras} with a less steep slope for the higher-energy GPs -- meaning that the wait-time for high-energy pulses is lower than would be expected based solely on extrapolating from the distribution of lower-energy GPs. Using over 2000\,hr of observations, \citet{kirsten_2024_natas} demonstrated a similar effect in the case of the hyperactive repeating FRB~20201124A. Future observations should observe M28A for $\sim1000$\,hr (using smaller and readily available telescopes), to search for a flattening of the GP energy distribution and to determine whether rare GPs from M28A can reach closer towards the energies observed for \frb. 

\begin{table*}
    \caption{Characteristics of GPs from M28A (this work) compared to bursts from \frb\ \citep{nimmo_2023_mnras}.}
    \label{tab:comparison_m28a_m81r}
    \small
    \setlength{\tabcolsep}{4pt}
    \begin{threeparttable}
    \centering
        \begin{tabular}{|p{0.2\linewidth}|p{0.25\linewidth}|p{0.25\linewidth}|p{0.11\linewidth}}
            \hline
            \hline
            \textbf{Observed characteristic}                     & \textbf{M28A GPs} (this work)                            & \textbf{\frb\ bursts} \citep{nimmo_2023_mnras} & \textbf{Link?}\\
            \hline
            Duration                                             & $\lesssim 2$\,\mus                                       & $50-100$\,\mus; variations as short as $\sim60$\,ns \citep{nimmo_2022_natas} & no\\
            \hline
            Spectral luminosities\tnote{a} & $\sim 10^{23}$\,erg~s\minusone~Hz\minusone               & $\sim 10^{28}$\,erg~s\minusone~Hz\minusone  & no \\ 
            \hline
            Energy distribution                                  & steep PL (index $\sim 1.8$)\tnote{b}                             & steep PL (index $\sim 2.4$) & yes \\
            \hline
            Wait-time distribution                               & Poissonian                                               & burst storms in which burst rate suddenly increases & no \\ 
            \hline
            Signs of periodicity                                 & highly periodic (occurrence in two narrow phase windows, but number of rotations between pulses random)  & no signs of periodicity & no \\
            \hline
            Frequency extent                                     & broad-band ($\sim 1-3$\,GHz) with sometimes narrow-band ($\sim 100$\,MHz) islands & narrow-band ($\lesssim$\,few~100~MHz) & some similarities \\
            \hline
            Burst spectra                                  & steep PL (index $\sim 1.4$) with Gaussian-peaked patches ($10-100$MHz) and sometimes isolated Gaussian-peaked islands ($\sim 100$\,MHz) & Gaussian peaked spectra & some similarities \\
            \hline
            DM variability                                       & $< 0.01$\,\dmunits\ on timescales of months     & $\lesssim 0.15$\,\dmunits\ between measurements separated by $\sim10$ months & yes \\
            \hline
        \end{tabular}
        \begin{tablenotes}
            \item\hspace{4.2mm} \textsuperscript{a} Isotropic equivalent.
            \item\hspace{4.2mm} \textsuperscript{b} Assuming constant pulse width and effective bandwidth between GPs.
        \end{tablenotes}
    \end{threeparttable}
\end{table*}

\section{Conclusions}
\label{sec:conclusions}
In this study, we investigated the hypothesis that \frb\ is an MSP producing extreme GPs. We conducted a detailed analysis of M28A (B1821$-$24A), the most energetic known MSP in our Galaxy (in terms of spin-down luminosity), and a well-known GP emitter \citep{knight_2006_apj,bilous_2015_apj}. 

In addition to considering previous studies, we performed new observations using the Parkes UWL receiver, with 2.0\,\mus\ time resolution and a $0.7-4.0$\,GHz frequency band. We identified $\sim 160$ GPs in 9\,hr of Parkes data. Our results provide new insights into the broad-band nature of M28A's GPs: using the 3.3-GHz observing bandwidth of the Parkes UWL receiver, our sub-banded search did not result in the detection of any genuinely narrow-band GPs ($< \text{few} 100$\,MHz). Nonetheless, we find that the GP spectra are sometimes composed of Gaussian-peaked $\lesssim 100$\,MHz narrow-band islands -- a frequency extent similar to what is observed in many repeating FRBs, including \frb.

We compared the observed characteristics of GPs from M28A with the bursts from \frb, as reported by \citet{nimmo_2023_mnras} -- the largest sample currently available. Our analysis revealed no strong links between the two sources, but we conclude that it would be valuable to perform high-cadence observations of M28A, for hundreds of hours using smaller radio dishes, to search for any rare but extremely luminous GPs that would be missed by our limited observing campaign. We observed the following distinctions and similarities in the characteristics of GPs from M28A compared to bursts from \frb:

\begin{enumerate}
    \item M28A's GPs have $\gtrsim 50\times$ shorter durations ($\lesssim 2$\,\mus) compared with \frb's bursts ($50-100$\,\mus); though \frb's bursts are atypically short duration compared to other repeating FRBs \citep{nimmo_2022_natas}, with the exception of the ultra-fast radio bursts found by \citet{snelders_2023_natas} from FRB~20121102A. Also, note that \frb\ has shown brightness variations {\it within bursts} on timescales as short as $\sim 60$\,ns \citep{nimmo_2022_natas}.
    
    \item M28A's GPs are 5 orders-of-magnitude lower in isotropic-equivalent spectral luminosity ($\sim 10^{23}$\,erg~s\minusone~Hz\minusone) compared with \frb's bursts ($\sim 10^{28}$\,erg~s\minusone~Hz\minusone); though the latter are atypically low-energy compared to other repeating FRBs \citep{nimmo_2022_natas}.
    
    \item M28A's GPs are strictly periodic, being emitted in two narrow phase windows, whereas no significant signs of periodicity are found for \frb\ \citep{nimmo_2023_mnras}. We note, however, that some neutron stars show both periodic pulses and brighter, aperiodic radio bursts \citep[e.g.][]{kirsten_2021_natas}.
    
    \item Using the S/N-distribution of our sample as a proxy for the energy distribution, the radio pulses from both sources are well described by a steep power-law energy distribution (with indices of $\sim 1.8$ and $\sim 2.4$ for M28A and \frb, respectively). We find no evidence for a flattening in the burst rate towards higher energies \citep[cf.][]{kirsten_2024_natas}, but a longer observing campaign is needed to probe this.
    
    \item In our observations, the occurrence of M28's GPs follow Poissonian statistics. In contrast, \frb\ can show burst storms, with a burst rate that changes drastically on timescales of minutes \citep{nimmo_2023_mnras}.
    
    \item M28A's GPs are generally broad-band ($\sim 1-3$\,GHz), but some show patchy structure composed of $\lesssim 100$\,MHz narrow-band islands. Wide-band detections of \frb\ are needed for better comparison, but some have observed frequency extents of $\lesssim$\,few~100\,MHz.

    \item M28A's GPs have a  steep spectral index $\sim 1.4$. Here too, broad-band studies of \frb\ are needed for better comparison.
    
    \item Both M28A and \frb\ show very little dispersion measure (DM) variability ($< 0.01$\,\dmunits) on timescales of months to years \citep{nimmo_2023_mnras}.
\end{enumerate}

\section*{Acknowledgements}
For the observations conducted with the Parkes' Murriyang telescope, we thank S. Dai, P. Kumar, G. Hobbs, and L. Toomey for their support.

The AstroFlash research group at McGill University, University of Amsterdam, ASTRON, and JIVE is supported by: a Canada Excellence Research Chair in Transient Astrophysics (CERC-2022-00009); the European Research Council (ERC) under the European Union’s Horizon 2020 research and innovation programme (`EuroFlash'; Grant agreement No. 101098079); and an NWO-Vici grant (`AstroFlash'; VI.C.192.045).

\section*{Data Availability}
The relevant code and data products from this work are publicly available via Zenodo: \url{https://doi.org/10.5281/zenodo.19283133}.

\bibliographystyle{mnras}
\bibliography{refs} 

%%%%%%%%%%%%%%%%% APPENDICES %%%%%%%%%%%%%%%%%%%%%
\appendix
\section{Additional context and analyses}

\begin{figure*}
  \centering
  \begin{minipage}{\textwidth} 
    \includegraphics[width=\textwidth]{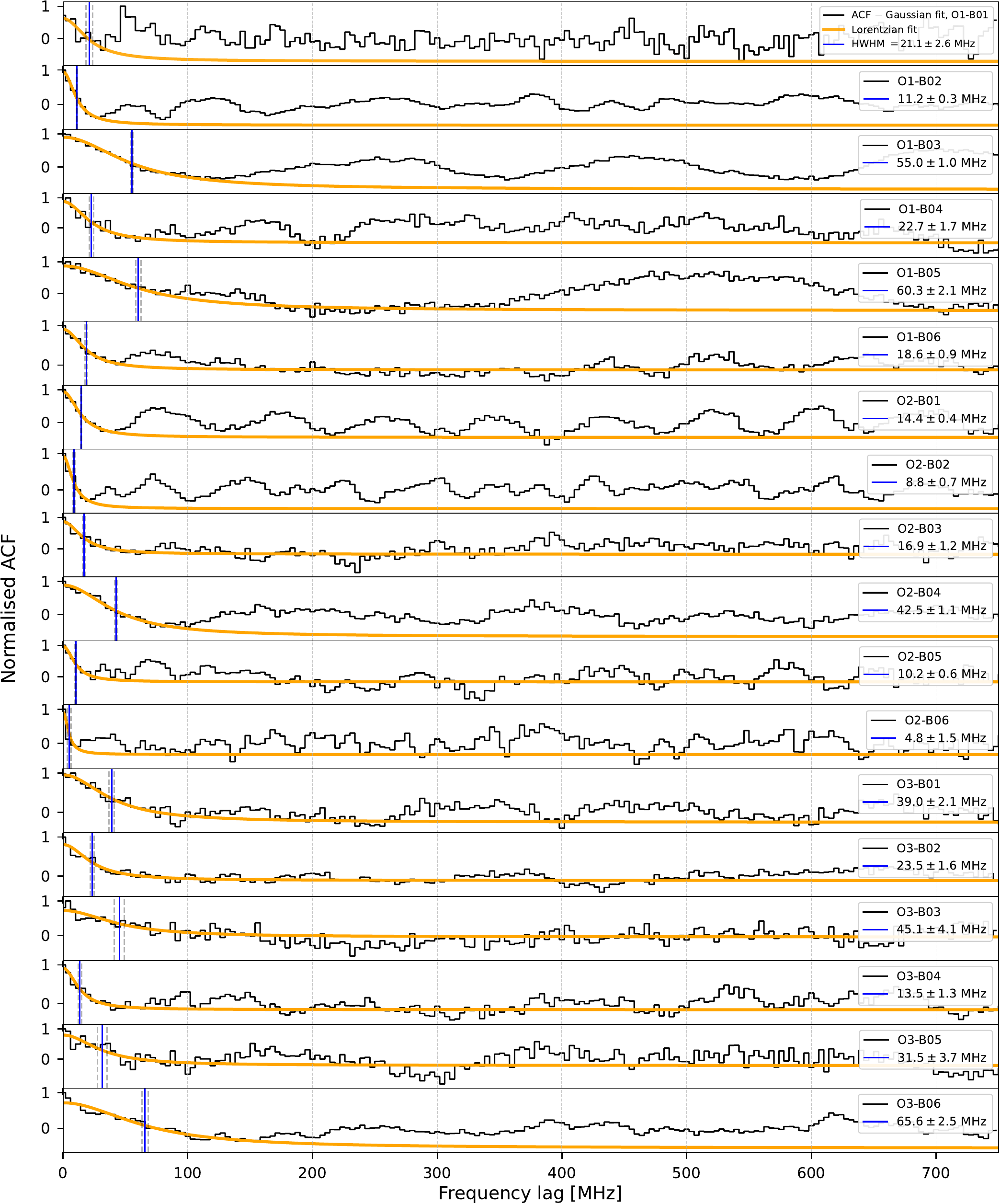}
  \end{minipage}
  \caption{The 1D ACFs for the 18 brightest GPs for the full band ($0.7-4.0$\,GHz) spectra (frequency resolution is $4$\,MHz, and the spectra are not smoothed with a Savitzky-Golay filter). The full-band ACF Gaussian shapes have been subtracted and the zero-frequency lag has been removed. The (ACF - Gaussian fit) residuals are represented by the black lines/steps, shown for up to $750$\,MHz frequency lag. For each ACF, a Lorentzian is fitted to the peak around zero, represented by orange lines.  The HWHM of de Lorentzian corresponds to the decorrelation bandwidth. The least-squares best-fit Lorentzian HWHM values, marked with vertical blue lines, are listed in the legends and range between between $4.8\pm1.5$\,MHz (O2-B06) and $65.6\pm2.6$\,MHz (O3-B06). The HWHM uncertainties are determined using the least-squares fit parameter covariance, and are represented by the vertical dashed light-blue lines.} 
  \label{fig:acf_O1_O2_O3_B01_to_B06}
\end{figure*}

\begin{figure}
	\includegraphics[width=\columnwidth]{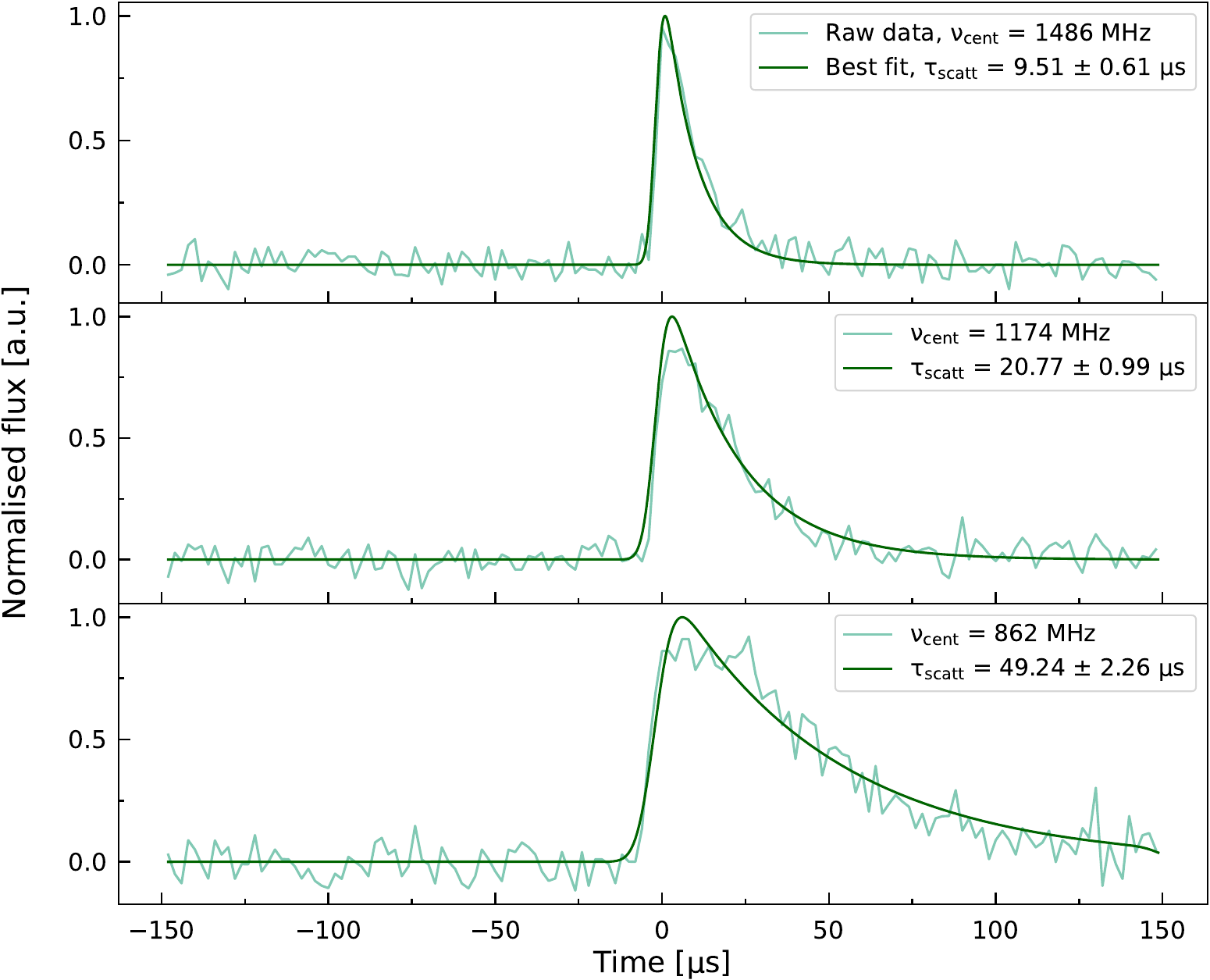}
    \caption{The total-intensity profile of the time-series of O1-B02 corresponding to $200$ MHz-wide sub-bands for a set of frequencies. Each profile is fitted with a Gaussian convolved with a one-sided exponential decay, to determine the scattering timescale $\tau_{\text{scatt}}$. The least-squares best-fit scattering timescales are $\tau_{\text{scatt}} = 9.51\pm0.61$\,\mus, $\tau_\text{scatt} = 20.77\pm0.99$\,\mus\ and $\tau_{\text{scatt}} = 49.24\pm2.26$\,\mus, corresponding to central frequencies $\nu_{\text{cent}} = 1486$\,MHz, $\nu_{\text{cent}} = 1174$\,MHz, and $\nu_{\text{cent}} = 862$\,MHz, respectively.}
    \label{fig:scattertail_fit_O1B02}
\end{figure}

\begin{figure}
	\includegraphics[width=\columnwidth]{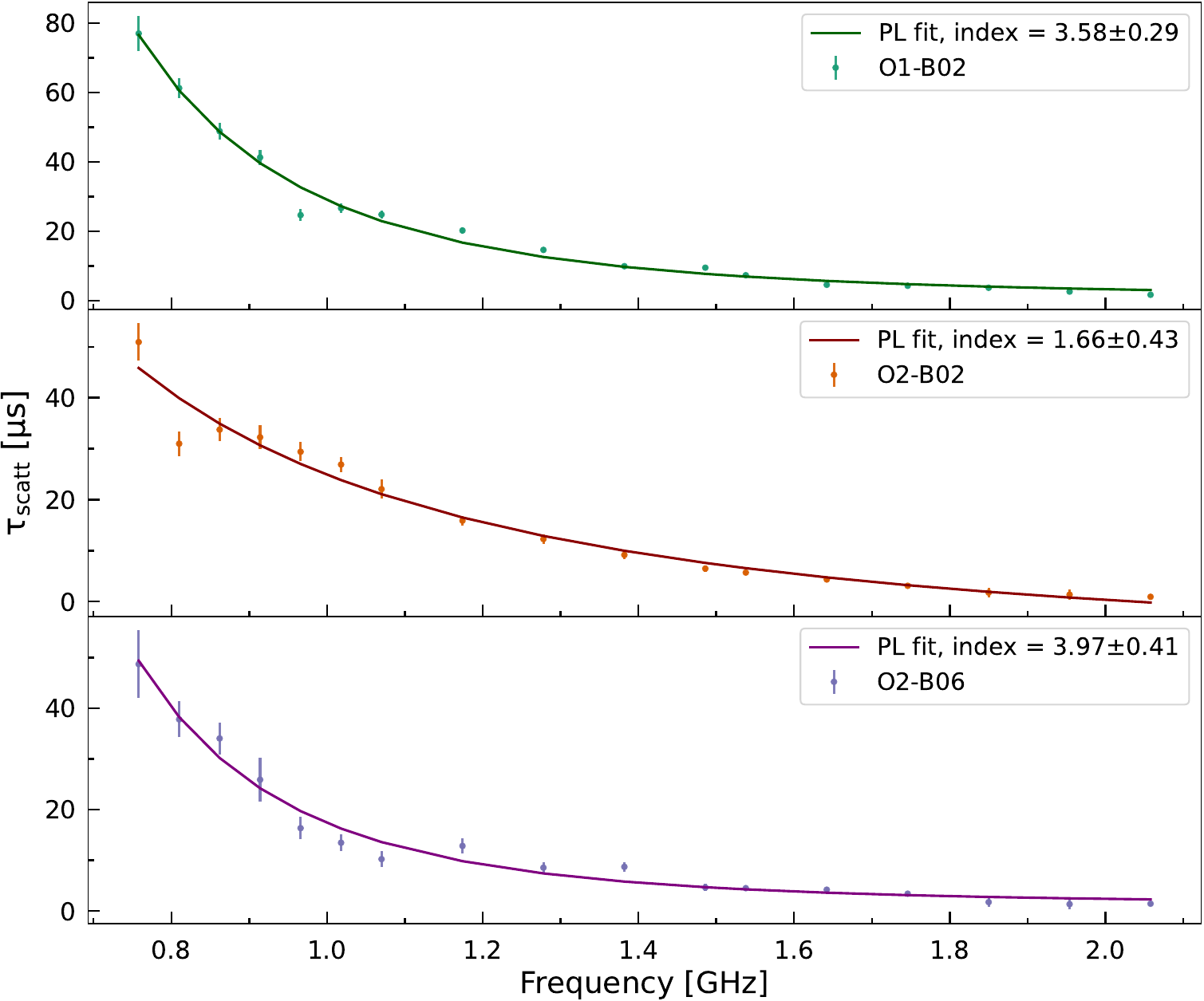}
    \caption{Measured scattering time scale versus frequency for the three brightest GPs, O1-B02 (upper panel, green), O2-B02 (middle panel, orange), and O2-B06 (lower panel, purple). Each data point represents the best-fit scattering timescale for a 200-MHz sub-band, see Figure\,\ref{fig:scattertail_fit_O1B02}. The least-squares best-fit power-law indices for O1-B02, O2-B02, and O2-B06 are $\beta = 3.58 \pm 0.29$, $\beta = 1.66 \pm 0.43$, and $\beta = 3.97 \pm 0.41$, respectively.}
    \label{fig:tscat_plfit_O1B02_O2B02_O2B06}
\end{figure}

\begin{figure*}
    \centering
    \includegraphics[width=\textwidth]{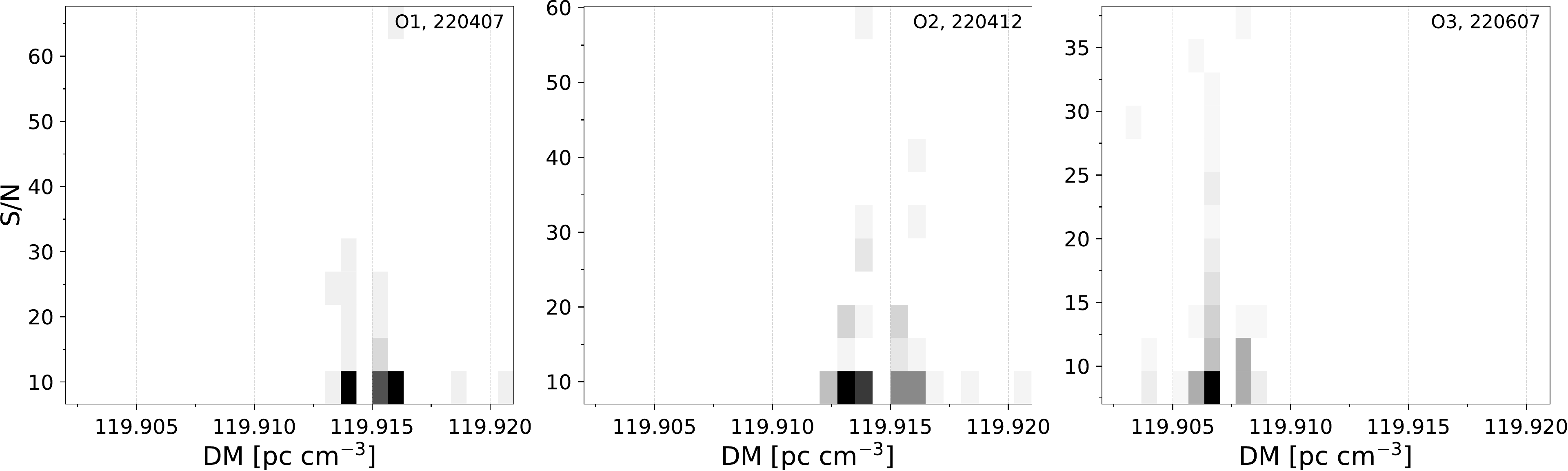}
    \caption{DM vs. S/N density distributions, showing maximised peak S/N and corresponding S/N-maximised DM determinations by \texttt{single\_pulse\_search.py} and \texttt{SpS.py}. The left panel shows the GP distribution for O1 (22-04-07), the middle panel for O2 (22-04-12), and the right panel for O3 (22-06-07). Darker shading indicates higher density (ranging from a single to about 10 bursts per bin). In two months, the average DM appears to have decreased by $\sim 0.01$\,\dmunits, from about $119.916$\,\dmunits\ in April 2022, to about $119.906$\,\dmunits\ in June 2022.}
    \label{fig:snrdm123}
\end{figure*}

\begin{figure}
    \centering
    \includegraphics[width=\columnwidth]{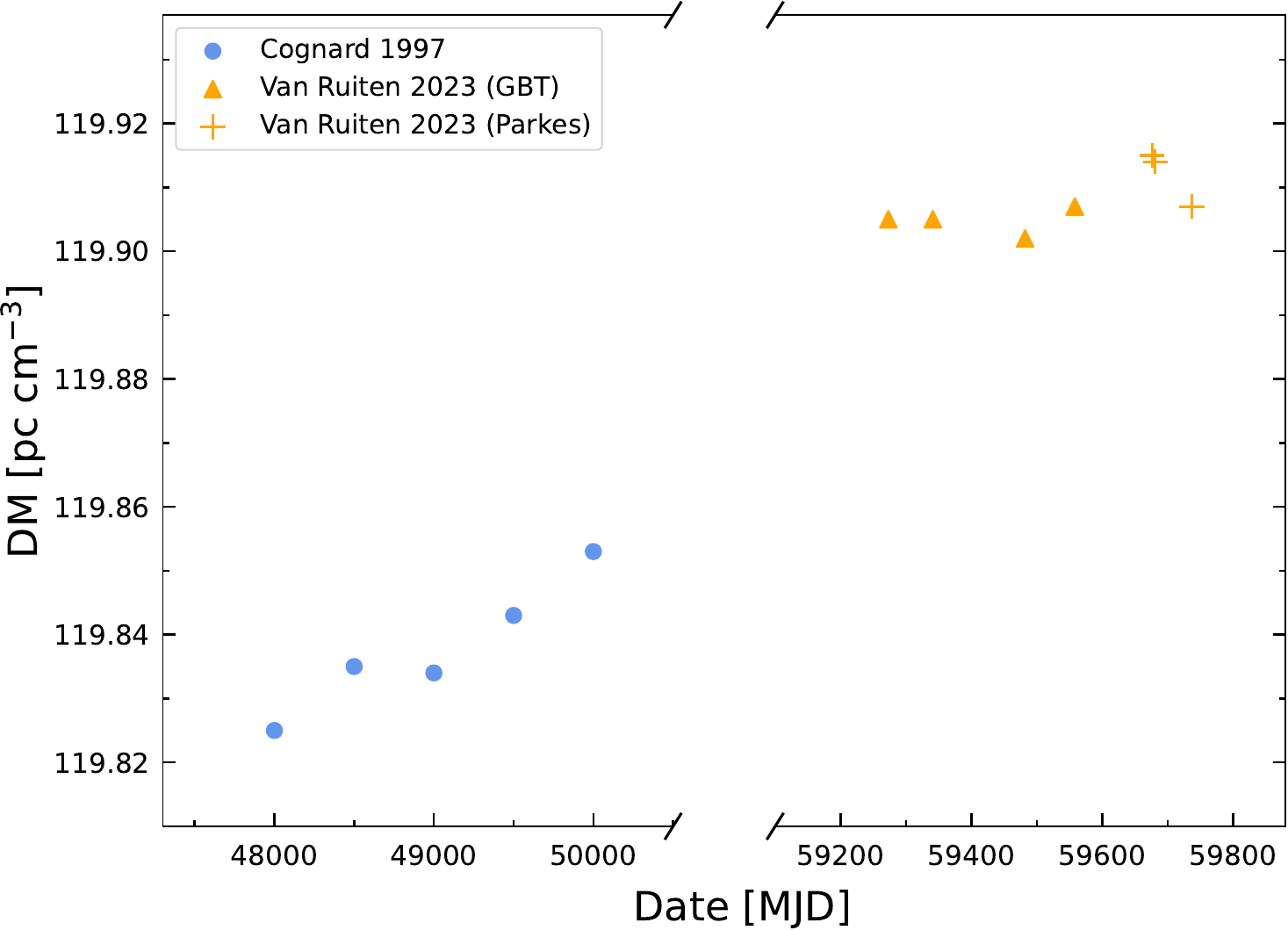}
    \caption{Diagnostic plot of the long-term variation in DM towards M28A observed in this research (represented by orange plus-signs; orange triangles represent our findings using M28A GPs we identified in archival GBT data) compared to observations conducted by \citet[][represented by blue dots, roughly averaged in this figure]{cognard_1997_aa}. The uncertainty on our DM measurements is $\sim 0.001$\,\dmunits, and is dominated by ambiguities introduced by frequency-dependent profile variations and scattering. On time-scales of months, the DM appears to be varying on scales of $\ge0.001$\,\dmunits. In $30$ years, the DM increased by $\sim 0.1$\,\dmunits, which translates roughly to an estimate of an average long-term DM increase of $\Delta \text{DM}/\text{d}t \sim 0.003$\,\dmunits~yr\minusone. Note the scale difference on the date-axis before and after the gap.}
    \label{fig:dm_longterm_var}
\end{figure}

%%%%%%%%%%%%%%%%%%%%%%%%%%%%%%%%%%%%%%%%%%%%%%%%%%
% do not change these lines
\bsp	% typesetting comment
\label{lastpage}
\end{document}